\documentclass[twocolumn,pra,floatfix,showpacs]{revtex4-1}
\usepackage{amsmath,amssymb,graphicx,epstopdf,epsfig}

\newcommand{\grad}{\boldsymbol{\nabla}}
\newcommand{\abs}[1]{\left|#1\right|}
\newcommand{\inlabs}[1]{|#1|}

\newcommand{\inleva}[1]{\langle#1\rangle}
\newcommand{\threevec}[3]{\left(\begin{array}{c}#1\\#2\\#3\end{array}\right)}
\newcommand{\nematic}{\mathbf{\hat{d}}}
\newcommand{\absF}{|\langle\mathbf{\hat{F}}\rangle|}
\newcommand{\inlabsF}{|\langle\mathbf{\hat{F}}\rangle|}
\newcommand{\expF}{\langle\mathbf{\hat{F}}\rangle}
\newcommand{\SO}{\mathrm{SO}}
\newcommand{\M}{\ensuremath{\mathcal{M}}}
\newcommand{\ket}[1]{\left|#1\right>}
\newcommand{\xhat}{\mathbf{\hat{x}}}
\newcommand{\yhat}{\mathbf{\hat{y}}}
\newcommand{\zhat}{\mathbf{\hat{z}}}
\newcommand{\rhohat}{\boldsymbol{\hat{\rho}}}
\newcommand{\phihat}{\boldsymbol{\hat{\varphi}}}
\newcommand{\Rtf}{R_\mathrm{TF}}
\newcommand{\F}{\mathbf{\hat{F}}}
\newcommand{\rr}{\mathbf{r}}
\newcommand{\nhat}{\mathbf{\hat{n}}}
\newcommand{\wno}[3]{\ensuremath{\left<#1,#2,#3\right>}}
\newcommand{\etaM}{\ensuremath{\eta_{\mathrm{f}}}}
\newcommand{\etaMs}{\ensuremath{\eta_{\mathrm{p}}}}
\newcommand{\etaMd}{\ensuremath{\eta_{\mathrm{d}}}}
\newcommand{\xin}{\ensuremath{\xi_n}}
\newcommand{\xiF}{\ensuremath{\xi_F}}
\newcommand{\xiZ}{\ensuremath{\xi_Z}}
\newcommand{\zetaN}{\ensuremath{\tilde{\zeta}}}
\newcommand{\NN}{\ensuremath{\tilde{N}}}
\newcommand{\MN}{\ensuremath{\tilde{M}}}
\newcommand{\MMN}{\ensuremath{\tilde{\M}}}
\newcommand{\betaN}{\ensuremath{\tilde{\beta}}}

\begin{document}

\author{Justin Lovegrove}
\author{Magnus O.\ Borgh}
\author{Janne Ruostekoski}
\affiliation{Mathematical Sciences, University of Southampton,
  Southampton, SO17 1BJ, United Kingdom}

\title{Stability and internal structure of vortices in spin-1
  Bose-Einstein condensates with conserved magnetization}

\begin{abstract}
We demonstrate how conservation of longitudinal magnetization can have
pronounced effects on both stability and structure of vortices in the atomic
spin-1 Bose-Einstein condensate by providing a systematic
characterization of nonsingular 
and singular vortex states.   Constructing spinor wave
functions for vortex states that continuously connect ferromagnetic and polar 
phases we systematically derive analytic models for nonrotating cores
of different singular vortices and for composite defect states with
distinct 
small- and large-distance topology. We explain how the conservation law
provides a stabilizing mechanism when the coreless vortex
imprinted on the condensate relaxes in the polar regime of interatomic
interactions. 
The resulting structure forms a composite defect: the
inner ferromagnetic coreless vortex deforms toward an outer singly
quantized polar vortex. We also numerically show how other even more complex 
hierarchies of vortex core topologies may be stabilized. Moreover, we
analyze the structure of the coreless vortex also in a ferromagnetic
condensate,  and show how reducing magnetization leads to a
displacement of the vortex from the trap center and eventually to the
deformation and splitting of its core where a singular vortex becomes
a lower-energy state. 
For the case of singular vortices, we find that the stability and the
core structure are notably less influenced by the conservation of
magnetization. 
\end{abstract}
\date{\today}
\maketitle

\section{Introduction}
\label{sec:introduction}

The multicomponent order parameters of spinor Bose-Einstein condensates
(BECs), created in all-optical traps~\cite{stamper-kurn_prl_1998} that
retain the atomic spin degree of freedom~\cite{stenger_nature_1998}, provide
a highly suitable laboratory for
the study of topological defects and textures. The accelerating interest in the
physics of defect structures in spinor BECs is illustrated by the very
recent realization of point
defects corresponding to analogs of Dirac magnetic
monopoles~\cite{ray_nature_2014} and 't~Hooft-Polyakov
monopoles~\cite{ray_science_2015}. In addition to the creation of
defects, there is increasing interest in the structure of the defect
cores as highlighted by the recent experimental observation of
the splitting of singly quantized vortices into pairs of half-quantum
vortices~\cite{seo_prl_2015}. These recent experiments follow on
experiments that have already
demonstrated controlled preparation of nonsingular
vortices and related
textures~\cite{leanhardt_prl_2003,leslie_prl_2009,choi_prl_2012,choi_njp_2012},
and out-of-equilibrium
production of singular vortices in rapid phase
transitions~\cite{sadler_nature_2006},
as well as observations of spontaneously formed spin
textures~\cite{vengalattore_prl_2008,kronjager_prl_2010}.
The experimental advances have been accompanied by a rapidly
increasing body of theoretical studies of spinor BECs (see e.g.\
Ref.~\cite{kawaguchi_physrep_2012} and 
references therein), describing, in particular,
singular and nonsingular vortices and
textures~\cite{yip_prl_1999,leonhardt_jetplett_2000,isoshima_pra_2002,mizushima_pra_2002,zhou_ijmpb_2003,saito_prl_2006,santos_prl_2006,semenoff_prl_2007,barnett_pra_2007,ji_prl_2008,kobayashi_prl_2009,huhtamaki_pra_2009,lovegrove_pra_2012,kobayashi_pra_2012,mawson_pra_2015,mizushima_prl_2002,martikainen_pra_2002,mueller_pra_2004,reijnders_pra_2004,takahashi_pra_2009,lovegrove_prl_2014} 
as well as
monopoles~\cite{stoof_prl_2001,ruostekoski_prl_2003,savage_pra_2003,pietila_prl_2009_dirac,ruokokoski_pra_2011}
and defect structures that cross a topological
interface~\cite{borgh_prl_2012,borgh_njp_2014}. Complex topological
textures are also possible in two-component (pseudospin-$1/2$)
BECs~\cite{ruostekoski_prl_2001,al-khawaja_nature_2001,battye_prl_2002,savage_prl_2003,kasamatsu_prl_2003,ruostekoski_pra_2004,takeuchi_jpsj_2006,kasamatsu_jhep_2010,mason_pra_2011,kawakami_prl_2012,nitta_pra_2012}.

When the energy of an optically trapped spinor BEC relaxes, its longitudinal
magnetization is approximately conserved on time scales where $s$-wave
scattering dominates the dynamics (over, e.g., weak dipolar interactions and
scattering with high-temperature
atoms; in the presence of resonant magnetic-field driving or spin-component
phase separation, the importance of the dipole-dipole interactions can
be
enhanced)~\cite{zhang_njp_2003,chang_prl_2004,makela_pra_2007,jacob_pra_2012}.
Here we provide a systematic characterization of the nonsingular
and singular vortex states in the spin-1 BEC,
and explain how conservation of
magnetization can have pronounced effects on their stability and
structure.
These are determined by analytic models and numerically by
relaxing the energy of suitable trial wave functions
representing the defect states.
To accurately account for conservation of magnetization, we
explicitly impose the constraint throughout the
energy-minimization procedure.

The spin-1 BEC exhibits two phases, ferromagnetic (FM) and polar,
depending on the spin-dependent part of the interaction, each with a
distinct family of vortices.
In Ref.~\cite{lovegrove_prl_2014} we
demonstrated that a FM coreless
vortex phase-imprinted on a spin-1 BEC with polar
interactions~\cite{leanhardt_prl_2003} can be energetically stable as
a composite topological defect when the conserved magnetization causes
the phases to mix. The
coreless vortex appears as the extended
core region in a defect structure 
that continuously approaches a singly
quantized polar vortex away from the vortex line.
Here we extend these studies by explicitly constructing spinor wave
functions for vortex states 
that interpolate smoothly between the FM and
polar phases. From the resulting analytic models we systematically
derive expressions for nonrotating cores of different singular
vortices, and a set of 
composite defect solutions with distinct small- and large-distance
topology. We then suggest how precise control over the magnitude 
and spatial profile of the quadratic Zeeman shift can be used to
stabilize also complex hierarchies of vortex core topologies.

Our analytic models for the vortex spinor wave functions 
that simultaneously describe singular vortices and their nonrotating superfluid
cores interpolate smoothly from the
ground state manifold to the orthogonal phase.  Whereas a vortex in a
scalar BEC always represents a depletion of the atomic density, such
filled cores can appear in the spinor BEC as energy relaxes as a
result of energetic competition in a hierarchy of characteristic
length scales.  The singularity of the order parameter is then
accommodated by exciting the spinor wave function out of the
ground-state manifold, becoming orthogonal to it at the singularity.
In Ref.~\cite{lovegrove_pra_2012} we demonstrated how this mechanism
leads to stable deformations of singly quantized vortices in both
phases of the spin-1 BEC.
We demonstrate here that these results are qualitatively robust when the
relaxation model is
refined to explicitly conserve magnetization as long as the
magnetization is not too strong, including the recently
observed~\cite{seo_prl_2015} splitting of a singly quantized polar
vortex into a pair of half-quantum vortices.  Our analytic solutions
then describe the relaxed vortex cores that could be observed in experiment.

Furthermore, we also provide a detailed analysis of a coreless vortex
in the FM regime when the magnetization is explicitly conserved. We
find that both the spin structure and the displacement of the 
coreless vortex are affected by the value of the magnetization:   
by displacing the vortex from the trap rotation axis, the total angular
momentum remains approximately constant for a range of magnetization
strengths. 
From our numerical results, we explain the mechanism by which a
coreless vortex becomes
energetically unstable when magnetization is weak, and gives way to a
singular-vortex ground state, as we briefly noted in
Ref.~\cite{lovegrove_prl_2014}.

The article is organized as follows:
In Secs.~\ref{sec:mft} 
and \ref{sec:numerics} 
we briefly review the mean-field theory of spin-1 BECs and outline the
numerical magnetization-conserving energy-minimization method.
Section~\ref{sec:topology-defects} 
gives an overview of the basic
vortices supported by the two phases. We then construct analytic
wave functions for vortex states that interpolate between the FM and
polar limits in Sec.~\ref{sec:interpolation}. 
Section~\ref{sec:coreless} 
investigates stability and structure of coreless
vortices in both FM and polar interaction regimes.
In Sec.~\ref{sec:weak-mag} 
we demonstrate the effects of a conserved weak magnetization on the
core structure of singular vortices.
We end by giving examples of how more complex core hierarchies can be
stabilized in a spin-1 BEC in section~\ref{sec:composite-core}, 
before summarizing our findings in section.~\ref{sec:conclusions}. 

\section{Spin-1 mean-field theory}
\label{sec:mft}

We treat the spin-1 BEC in the
Gross-Pitaevskii mean-field theory.
The Hamiltonian density then
reads~\cite{kawaguchi_physrep_2012,stamper-kurn_rmp_2013}
\begin{equation}
  \label{eq:hamiltonian-density}
  \begin{split}
    \mathcal{H} &=  \frac{\hbar^2}{2m}\abs{\grad\Psi}^2 +
    \frac{1}{2}m\omega^2r^2n
    + \frac{c_0}{2}n^2
    + \frac{c_2}{2}n^2|\mathbf{\inleva{\hat{F}}}|^2\\
    &+ g_1n\inleva{\mathbf{B}\cdot\mathbf{\hat{F}}}
    + g_2n\inleva{(\mathbf{B}\cdot\mathbf{\hat{F}})^2},
  \end{split}
\end{equation}
where $n=\Psi^\dagger\Psi$ is the atom density and $m$ is the atomic
mass. We have
assumed an isotropic, harmonic trap of frequency $\omega$.
The condensate wave function $\Psi$ is now a three-component spinor,
\begin{equation}
  \label{eq:spinor_suppl}
  \Psi({\bf r}) = \sqrt{n({\bf r})}\zeta({\bf r})
  = \sqrt{n({\bf r})}\threevec{\zeta_+({\bf r})}
                              {\zeta_0({\bf r})}
                              {\zeta_{-}({\bf r})},
  \quad
  \zeta^\dagger\zeta=1,
\end{equation}
in the basis of spin projection onto the $z$ axis. The local
condensate spin is the expectation value
$\inleva{\F}=\zeta_\alpha^\dagger\mathbf{\hat{F}}_{\alpha\beta}\zeta_\beta$
of the spin operator $\F$, defined as a vector of spin-1 Pauli
matrices.
Linear and quadratic Zeeman shifts, of strength $g_1$ and $g_2$
respectively,
described by the last two terms of Eq.~\eqref{eq:hamiltonian-density}, may
arise from a weak external magnetic field $\mathbf{B}$.

Spin-independent and spin-dependent interaction terms with
strengths $c_0=4\pi\hbar^2(2a_2+a_0)/3m$ and
$c_2=4\pi\hbar^2(a_2-a_0)/3m$, respectively, arise from the two
scattering channels of colliding spin-1 atoms with $s$-wave scattering
lengths $a_0$ and $a_2$.
Minimization of the interaction energy then
leads to the two distinct phases of the spin-1 BEC: $c_2<0$ favors the
$\absF=1$ ferromagnetic (FM) phase (e.g., in $^{87}$Rb), while the $\absF=0$
polar phase is favored when $c_2>0$ (e.g., in $^{23}$Na).

We find stable vortex structures by minimizing the free energy
$E=\int d^3r\,\mathcal{H} - \Omega\inleva{\hat{L}_z}$
in the frame rotating at frequency
$\Omega$ about the $z$ axis, using imaginary-time propagation of the
coupled Gross-Pitaevskii equations.
However, the only spin-flip
processes possible in $s$-wave scattering are
\mbox{$2\ket{m=0} \rightleftharpoons \ket{m=+1} + \ket{m=-1}$.} Therefore
$s$-wave interaction
does not change
the \emph{longitudinal magnetization}
\begin{equation}
  \label{eq:M}
    M = f_+-f_- = \frac{1}{N}\int d^3r\, n(\rr)F_z(\rr)\,,
\end{equation}
where $f_\pm= N_\pm/N$. Here the total and the $\ket{m=\pm1}$ level populations
are denoted by $N$ and $N_\pm$, respectively. We have also introduced
the $z$-component of the spin $F_z=\bf{\hat z}\cdot\inleva{\F}$.

Consequently, $M$ is approximately
conserved on time scales where $s$-wave scattering
dominates over, e.g., dipolar interactions and collisions with
high-temperature atoms. This is the relevant time scale in present
experiments with spinor BECs of alkali-metal
atoms~\cite{stenger_nature_1998,chang_prl_2004,jacob_pra_2012}.
We take this conservation strictly into account throughout energy
relaxation by
simultaneously renormalizing $N$ \emph{and} $M$ at each step of
imaginary-time evolution.

The interaction terms in Eq.~(\ref{eq:hamiltonian-density}) give rise
to the characteristic density and spin healing lengths,
\begin{equation}
\label{eq:healinglengths}
  \xin = l\left(\frac{\hbar\omega}{2 c_0 n}\right)^{1/2}, \quad
  \xiF = l\left(\frac{\hbar\omega}{2 |c_2| n}\right)^{1/2},
\end{equation}
where we have introduced the oscillator length
$l=(\hbar/m\omega)^{1/2}$ of the harmonic confinement.
The healing lengths determine, respectively, the length scales
over which the atomic
density $n(\rr)$ and the spin magnitude $\absF$
heal around a local
perturbation.
When magnetization is not conserved, $\xin$ and $\xiF$ determine
the core size of singular
defects~\cite{ruostekoski_prl_2003,lovegrove_pra_2012}. If
$\xiF$ is sufficiently
larger than $\xin$, it becomes energetically favorable to avoid
depleting the atomic density, instead accommodating the singularity
by exciting the wave function out of its ground-state manifold. The
core then expands to the order of $\xiF$, instead of the smaller
$\xin$ that determines the size of a core with vanishing density. The
lower gradient energy in the larger core offsets the cost in
interaction energy.

Conservation of magnetization introduces a third length scale
$\etaM$, which is the size required for a magnetized vortex core
in an otherwise unmagnetized condensate to give rise to a given
magnetization. In order to estimate the magnetization length scale
we represent the magnetized core by a cylinder of
radius $\etaM$, with $\expF=\zhat$ everywhere inside
the core and $\absF=0$ outside. The total magnetization is then
\begin{equation}
\label{eq:cyl-mag}
  M(\etaM) = \frac{1}{N}\int d^3r\,
  \Theta(\etaM-\rho) n_{\rm TF}(\mathbf{r})\,,
\end{equation}
where $\rho=(x^2+y^2)^{1/2}$ and $\Theta$ is the Heaviside function.
We approximate the atomic-density profile by the Thomas-Fermi solution
\begin{equation}
  n_{\rm TF}(r)=\frac{15N}{8\pi
  \Rtf^3}\left(1-\frac{r^2}{\Rtf^2}\right),
  \quad r\leq\Rtf\,,
\end{equation}
where $r=(\rho^2+z^2)^{1/2}$, and
\begin{equation}
  \Rtf
    = l\left(\frac{15}{4\pi}\frac{Nc_{\rm p,f}}{\hbar\omega l^3}\right)^{1/5}
\end{equation}
is the
Thomas-Fermi radius. Here $c_{\rm p}=c_0$ in a BEC with polar interactions,
and $c_{\rm f}=c_0+c_2$ in the FM regime.
Computing the integral in Eq.~\eqref{eq:cyl-mag} and solving for
$\etaM$ as a function of $M$, we obtain
\begin{equation}
  \label{eq:magnetization-healing-length}
  \etaM = \Rtf\sqrt{1-\left(1-M\right)^{2/5}}\,.
\end{equation}

An analogous length scale $\etaMs$ may be defined
as the size required for an unmagnetized vortex core
in an otherwise magnetized condensate to give rise to a given
magnetization. We represent the unmagnetized core by a cylinder
of radius $\etaMs$, with $\absF=0$ everywhere inside
the core and $\expF=\zhat$ outside. The total magnetization
may then be calculated, again by assuming a Thomas-Fermi
density profile, as
\begin{equation}
  M(\etaMs) = \frac{1}{N}\int d^3r\,
  \Theta(\rho-\etaMs) n_{\rm TF}(\mathbf{r})\,.
\end{equation}
Solving for $\etaMs$ yields
\begin{equation}
  \etaMs = \Rtf\sqrt{1-M^{2/5}}\,.
\label{eq:maglength-polcore}
\end{equation}

\section{Numerical method}
\label{sec:numerics}

The energetic stability of vortex configurations, and the corresponding
stable core structures, can be determined by numerically minimizing
the free energy of suitable initial spinor wave functions in the rotating
frame. Spinor wave functions representing specific
vortices that can be used as such prototypes are constructed in
Sec.~\ref{sec:topology-defects}. We minimize the energy by
propagating the coupled Gross-Pitaevskii
equations derived from Eq.~(\ref{eq:hamiltonian-density}) in imaginary
time. Due to the renormalization of the wave function in each time
step, the longitudinal magnetization $M$
[Eq.~(\ref{eq:M})], can change during the energy relaxation.
However, the interaction terms in
Eq.~(\ref{eq:hamiltonian-density}) arise from $s$-wave scattering,
which preserves the relative populations of the $\zeta_\pm$ spinor
components. Therefore $M$ is approximately conserved on
the time scales of interest here, where $s$-wave scattering dominates
and we may neglect processes that cause magnetization to relax (e.g.,
due to dipolar interactions).

In order to conserve $M$ throughout the
relaxation process, the imaginary-time propagation is performed using
a split-step algorithm~\cite{javanainen_jpa_2006}, where both the
total number of atoms $N$ and the magnetization $M$ are
constrained in each time step. This represents a different physical
mechanism than that used in numerical techniques where an effective linear
Zeeman term is used as a Lagrange multiplier for
the magnetization of the final state, without explicitly conserving
magnetization during relaxation~\cite{yip_prl_1999,isoshima_pra_2002,mizushima_prl_2002,mizushima_pra_2002}.
With our method, the value of the
conserved magnetization is determined by the initial state. We
therefore construct initial wave functions from the
appropriate spinors in
Sec.~\ref{sec:topology-defects} by adjusting the relative populations
of the components to yield the desired magnetization.

We consider
a condensate in an isotropic, three-dimensional (3D), harmonic trap
with
interaction strengths chosen such that $Nc_0=1000\hbar\omega l^3$.
The spin-dependent nonlinearity is kept fixed at $Nc_2=-5\hbar\omega
l^3$ ($Nc_2=36\hbar\omega l^3$) in the FM (polar) regime,
which is consistent with the experimentally measured ratio of
$c_2/c_0$ for $F=1$ $^{87}$Rb
($^{23}$Na)~\cite{van-kempen_prl_2002,knoop_pra_2011}.

\section{Basic vortices in the spin-1 BEC}
\label{sec:topology-defects}

The set of physically distinguishable, energetically degenerate states
defines the order-parameter manifold, the symmetry properties of which
determine
the families of vortices it supports.
Here we
give a brief overview of the basic spin-1 vortices in both phases, before
discussing the effects of magnetization in
sections~\ref{sec:coreless} and \ref{sec:weak-mag}. A more complete
presentation can be found for example in
Refs.~\cite{lovegrove_pra_2012,borgh_pra_2013}.

\subsection{Vortices in the  FM phase}
\label{sec:fm-vortices}

The atom-atom interaction in the spin-1 BEC is in the FM regime if the
sign of the spin-dependent contribution is negative,
such that $c_2<0$ in Eq.~(\ref{eq:hamiltonian-density}). Then the
interaction strives to maximize the magnitude of the spin: $\absF=1$
everywhere when the spin texture is uniform. We can write such a state
simply as $\zeta^{(1)} = (-1,0,0)^T$, such that the spin vector is parallel
to the $z$ axis. (The physically insignificant overall minus
  sign in $\zeta^{(1)}$ is included for later convenience.) Any other FM
spinor can then be
reached by a spin rotation, defined by three Euler
angles $\alpha$,
$\beta$ and $\gamma$, and a condensate phase $\tau$, yielding
\begin{equation}
  \label{eq:ferro}
  \zeta^{\rm f}
  = \frac{-e^{i(\tau-\gamma)}}{\sqrt{2}}
    \threevec{\sqrt{2}e^{-i\alpha}\cos^2\frac{\beta}{2}}
             {\sin\beta}
             {\sqrt{2}e^{i\alpha}\sin^2\frac{\beta}{2}}.
\end{equation}
The third Euler angle and the condensate phase can be combined into
$\tau^\prime=\tau-\gamma$.
The FM order-parameter manifold
is then the group of rotations in three dimensions, $\SO(3)$, and
the Euler angles $\alpha$ and $\beta$ give the local expectation value
of the spin vector
\begin{equation}
  \label{eq:FMexpF}
  \expF=\cos\alpha\sin\beta \mathbf{\hat{x}}
        + \sin \alpha\sin\beta \mathbf{\hat{y}} + \cos\beta\zhat.
\end{equation}

We can classify vortices by considering how the order parameter
changes on a loop around the vortex line~\cite{mermin_rmp_1979}.
For the FM order-parameter space $\SO(3)$ this gives two types of
vortices: nonsingular and singular, singly quantized vortices.
As a consequence of the expanded broken symmetry of the spinor wave function,
the mass circulation alone is not necessarily quantized in the
FM phase. Since the
superfluid velocity in the FM phase is expressed in terms of $\alpha$,
$\beta$ and $\tau^\prime$ as
\begin{equation}
  \label{eq:sfv}
  \mathbf{v} = \frac{\hbar}{m}
  \left(\mathbf{\nabla}\tau^\prime-\cos\beta\mathbf{\nabla}\alpha\right),
\end{equation}
a continuous variation in $\beta$ enables the circulation, $\nu$, to take a
value anywhere in the continuous interval
$\oint\mathbf{\nabla}(\tau^\prime-\alpha)\cdot d\mathbf{r} \leq \nu
\leq \oint\mathbf{\nabla}(\tau^\prime+\alpha)\cdot d\mathbf{r}$.
In the lowest-energy
configurations of both the singular and the nonsingular FM vortices,
$\beta$ varies spatially~\cite{lovegrove_pra_2012},
such that the circulation is in general not quantized (except when a
magnetization $M=\pm1$ forces the
condensate to the one-component limit).

A nontrivial, nonsingular vortex---a coreless vortex---may be
constructed by combining a $2\pi$ rotation of
the spin vector with a simultaneous $2\pi$ winding of the condensate
phase, corresponding to the choice
$\alpha=\tau^\prime=\varphi$, where $\varphi$ is the azimuthal
angle, in Eq.~(\ref{eq:ferro}):
\begin{equation}
  \label{eq:cl}
  \zeta^{\rm cl}(\mathbf{r}) =
  \frac{-1}{\sqrt{2}}\threevec{\sqrt{2}\cos^2\frac{\beta(\rho)}{2}}
                          {e^{i\varphi}\sin\beta(\rho)}
                          {\sqrt{2}e^{2i\varphi}\sin^2\frac{\beta(\rho)}{2}}.
\end{equation}
Here $\beta$ describes how the spin vector tilts away
from the $z$ axis as the radial distance $\rho$ increases. This results in a
characteristic fountain-like spin texture
\begin{equation}
  \label{eq:fm-fountain}
  \inleva{\F}=\sin\beta\rhohat + \cos\beta\zhat
\end{equation}
that is continuous
everywhere, with $\inleva{\mathbf{\hat{F}}}=\mathbf{\hat{z}}$ on the
vortex line. 

The coreless vortex is similar to the Anderson-Toulouse-Chechetkin
and Mermin-Ho vortices in superfluid
$^3$He~\cite{chechetkin_jetp_1976,anderson_prl_1977,mermin_prl_1976},
where a circulation-carrying nonsingular texture is formed by the
angular-momentum vector $\mathbf{l}$ of the cooper pairs. The two
vortices have different asymptotic $\mathbf{l}$-texture
away from the vortex line: antiparallel to the vortex line in the
Anderson-Toulouse-Chechetkin 
texture, and  perpendicular to it in the Mermin-Ho texture.

Owing to the effectively two-dimensional (2D) structure of the
coreless spin texture, it is possible to define a winding number 
\begin{equation}
\label{eq:coreless-charge}
  W= \frac{1}{8\pi} \int_\mathcal{S} {\rm d}\Omega_i \epsilon_{ijk}
  \nhat_F \cdot
  \left(\frac{\partial\nhat_F}{\partial x_j} \times
  \frac{\partial\nhat_F}{\partial x_k}\right)\,.
\end{equation}
Here the integral is evaluated over a surface $\mathcal{S}$ that
covers the full cross-section of the coreless-vortex texture.
We have defined $\nhat_F=\inleva{\F}/\absF$ as a unit vector in the
direction of the local spin vector, in order to later consider
coreless textures where the condensate is not everywhere strictly in
the FM phase.
The charge $W$ defines a topological invariant if the boundary
condition on $\nhat_F$ away from the vortex is fixed,
e.g., by physical interaction or
energetics.
When the asymptotic texture is uniform, $W$ is an integer
(representing a mapping of the spin texture on a compactified
2D plane onto the unit sphere).
If no boundary condition is imposed, the texture can
unwind to the vortex-free state by purely local transformations of the
wave function.
As a result of \eqref{eq:coreless-charge}, coreless textures are
also called 2D ``baby Skyrmions''~\cite{manton-sutcliffe} in analogy
with the 3D Skyrmions~\cite{skyrme_1961}, which represent stable
particlelike solitons that can also exist in atomic
BECs~\cite{ruostekoski_prl_2001,battye_prl_2002,savage_prl_2003,ruostekoski_pra_2004,kawakami_prl_2012,nitta_pra_2012}.

The spin-1 coreless vortex may be stabilized by
rotation as the bending angle $\beta(\rho)$ in Eq.~\eqref{eq:cl}, and
therefore the superfluid circulation, adapts to minimize the energy.
The asymptotic behavior of the spin texture is then determined
by the imposed rotation.
The spin texture away from the vortex line
may also be determined by interactions with other vortices, e.g., in the
formation of a composite defect.
By finding $\expF$ from Eqs.~\eqref{eq:cl} and \eqref{eq:fm-fountain}, and
substituting in Eq.~\eqref{eq:coreless-charge}, we may evaluate $W$
assuming cylindrical symmetry.  Taking $R$ to be the radial extent of the
spin texture, we find
\begin{equation}
  \label{eq:cl-W}
  W = \frac{1-\cos\beta(R)}{2}\,,
\end{equation}
where we have used $\beta=0$ on the $z$ axis, such
that $\left.\nhat_F\right|_{\rho=0}=\zhat$. The winding number now
depends on the asymptotic value
of $\beta(\rho)$, such that for $\beta(R)=\pi$ (Anderson-Toulouse-Chechetkin-like texture)
$W=1$, and for $\beta(R)=\pi/2$ (Mermin-Ho-like texture) $W=1/2$.

The only other class of vortices in the FM phase is formed by the
singly quantized vortices.
We construct a simple representative as a $2\pi$ winding of the
condensate phase in a uniform spin texture, such that
$\tau^\prime=\varphi$ in Eq.~(\ref{eq:ferro}), leaving $\alpha=\alpha_0$ and
$\beta=\beta_0$ constant:
\begin{equation}
  \label{eq:fmsingular}
  \zeta^{\rm s} =
  \frac{-e^{i\varphi}}{\sqrt{2}}
  \threevec{\sqrt{2}e^{-i\alpha_0}\cos^2\frac{\beta_0}{2}}
           {\sin\beta_0}
           {\sqrt{2}e^{i\alpha_0}\sin^2\frac{\beta_0}{2}}.
\end{equation}
Vortices in the same equivalence class can be transformed into each
other by local spin rotations.
For example,
$\inleva{\mathbf{\hat{F}}}$ may form a radial disgyration, 
with $\alpha=\varphi$,
$\tau'=0$ and constant $\beta=\beta_0$, represented by
\begin{equation}
  \label{eq:spinvortex}
  \zeta^{\rm sv} =
  \frac{-1}{\sqrt{2}}\threevec{\sqrt{2} e^{-i\varphi} \cos^2\frac{\beta_0}{2}}
                      {\sin\beta_0}
                      {\sqrt{2} e^{i\varphi} \sin^2\frac{\beta_0}{2}}.
\end{equation}
For $\beta_0=\pi/2$ the spins lie in the $(x,y)$ plane, forming a spin
vortex.
Local spin rotations in a singular vortex state may occur, for example,
as a result
of energy relaxation of the vortex core~\cite{lovegrove_pra_2012}.

The spatially varying spin texture of the relaxed vortex state determines the
longitudinal magnetization of the
condensate. It is therefore a nontrivial question how
  conservation of a given initial magnetization in an experimentally
  prepared vortex state affects the spin texture and vortex core
  structure of relaxed state.

\subsection{Polar phase}
\label{sec:polar-vortices}

The polar phase minimizes the magnitude of the spin, $\absF=0$
everywhere in a
uniform spin texture, and is energetically preferred when $c_2>0$ in
Eq.~(\ref{eq:hamiltonian-density}). Similarly to the description
  of the FM phase, we find an expression
  for a general polar state by applying a spin rotation and a
  condensate phase to a representative polar spinor, which we take to
  be $\zeta^{(0)}=(-1/\sqrt{2},0,1/\sqrt{2})^T$. The reason for this
  particular choice will become clear in Sec.~\ref{sec:interpolation}.
In terms of the condensate phase and the three Euler angles defining
the spin rotation, the general polar spinor is then
\begin{equation}
  \label{eq:polar}
  \zeta^{\rm p}
  = \frac{e^{i\tau}}{\sqrt{2}}\threevec{ e^{-i \alpha}\left(e^{ i \gamma} \sin^2\frac{\beta}{2}-e^{-i \gamma}\cos^2\frac{\beta}{2}\right)}
{-\sqrt{2}\sin\beta\cos \gamma}
{e^{i \alpha}\left(e^{i  \gamma} \cos^2\frac{\beta}{2}- e^{-i \gamma}\sin^2\frac{\beta}{2}\right)}.
\end{equation}

Note that $e^{-i\tau}\zeta_+=-e^{i\tau}\zeta_-^*$. 
Equation~\eqref{eq:polar} can
therefore be written in the form~\cite{ruostekoski_prl_2003}
\begin{equation}
\label{eq:nematic}
  \zeta^{\rm p} = \frac{e^{i\tau}}{\sqrt{2}}
                 \threevec{-d_x+id_y}{\sqrt{2}d_z}{d_x+id_y},
\end{equation}
where $\nematic=d_x\xhat+d_y\yhat+d_z\zhat$ is a unit vector.
In our
choice of representative spinor $\zeta^{(0)}$, $\nematic=\xhat$, and it
follows that in the general polar spinor,
$\nematic=(\cos\alpha \cos\beta \cos\gamma - \sin\alpha
\sin\gamma)\mathbf{\hat{x}}+
(\sin\alpha \cos\beta \cos\gamma + \cos\alpha \sin\gamma)\mathbf{\hat{y}}-
\cos\gamma \sin\beta\mathbf{\hat{z}}$
in terms of the Euler angles.
Two angles (polar and azimuthal) are sufficient to give the
direction of a vector. However, anticipating the construction of
phase-mixing vortex states in Sec.~\ref{sec:interpolation}, we choose
to retain the three Euler angles, keeping in mind that more than one
combination may yield the same $\nematic$.
Note that
$\zeta^{\rm p}(\tau,\nematic) = \zeta^{\rm p}(\tau+\pi,-\nematic)$.
This property is called
nematic order~\cite{zhou_ijmpb_2003}, and the vector $\nematic$,
called the nematic axis, is understood as unoriented.

From Eq.~(\ref{eq:polar}) one can show that the superfluid
velocity in the polar phase is given by
\begin{equation}
  \label{eq:polar-velocity}
  \mathbf{v} = \frac{\hbar}{m}\grad\tau.
\end{equation}
It then follows by the standard argument (see, e.g.,
Ref.~\cite{kawaguchi_physrep_2012}) that circulation must be quantized.
As in the FM phase, a singly quantized vortex can
be constructed as a $2\pi$ winding
of the condensate phase: $\tau=\varphi$ in Eq.~(\ref{eq:polar}) and
keeping $\alpha=\alpha_0$, $\beta=\beta_0$ and
$\gamma=\gamma_0$ constant to form
\begin{equation}
  \label{eq:psingular}
  \zeta^{\rm 1q} =
  \frac{e^{i\varphi}}{\sqrt{2}}\threevec{ e^{-i \alpha_0}\left(e^{ i \gamma_0} \sin^2\frac{\beta_0}{2}-e^{-i \gamma_0}\cos^2\frac{\beta_0}{2}\right)}
{-\sqrt{2}\sin\beta_0\cos \gamma_0}
{e^{i \alpha_0}\left(e^{i  \gamma_0} \cos^2\frac{\beta_0}{2}- e^{-i \gamma_0}\sin^2\frac{\beta_0}{2}\right)}
\end{equation}
However, because the superfluid velocity depends only on the
condensate phase $\tau$, a singly quantized vortex
may also include, e.g., a $2\pi$ winding of $\alpha$
without changing the circulation.
The vortex is then represented by
\begin{equation}
  \label{eq:p012}
  \zeta^{\rm 1q^\prime} =
  \frac{1}{\sqrt{2}}\threevec{ \left(e^{ i \gamma_0} \sin^2\frac{\beta_0}{2}-e^{-i \gamma_0}\cos^2\frac{\beta_0}{2}\right)}
{-\sqrt{2}e^{i\varphi}\sin\beta_0\cos \gamma_0}
{e^{i2\varphi}\left(e^{i  \gamma_0} \cos^2\frac{\beta_0}{2}- e^{-i \gamma_0}\sin^2\frac{\beta_0}{2}\right)},
\end{equation}
keeping $\beta=\beta_0$ and $\gamma=\gamma_0$ constant.

Unlike a scalar BEC, the singly quantized vortex in the polar spin-1
BEC does not represent the smallest unit of circulation.
The nematic order
makes it possible to form a vortex carrying half a quantum
of circulation by allowing the condensate phase to wind by only $\pi$
as the vortex line is encircled. A simultaneous spin rotation of $\nematic$
into $-\nematic$ keeps the order parameter single-valued.
If the vortex line is on the $z$ axis, and $\nematic$ is in the
$(x,y)$ plane, this corresponds to choosing $\tau=\alpha=\varphi/2$
and $\beta=\pi/2$ in
Eq.~(\ref{eq:polar}), such that
\begin{equation}
  \label{eq:hq}
  \zeta^{\rm hq} = \frac{e^{i\varphi/2}}{\sqrt{2}}
                   \threevec{-e^{-i\varphi/2}}
                            {0}
                            {e^{i\varphi/2}}
		 = \frac{1}{\sqrt{2}}
                   \threevec{-1}
                            {0}
                            {e^{i\varphi}}.
\end{equation}
Similar
half-quantum vortices arise from the nematic order in superfluid
$^3$He-$A$~\cite{salomaa_prl_1985,salomaa_rmp_1987}, and are also
found in nematic liquid crystals~\cite{degennes-prost}.

The quantization of circulation in the polar phase prevents
angular-momentum carrying nonsingular vortices from
forming. Nevertheless it is
possible for the $\nematic$ vector to form a nonsingular texture
similar to the spin texture of the coreless vortex. It does not,
however, carry mass circulation. Such a \emph{nematic coreless vortex}
is constructed as a $2\pi$
spin rotation, corresponding to $\alpha=\varphi$ and  $\gamma=\pi$ in
Eq.~\eqref{eq:polar}:
\begin{equation}
  \label{eq:ncv}
  \zeta^{\rm ncv}
  = \frac{1}{\sqrt{2}}\threevec{ e^{-i \varphi}\cos\beta(\rho)}
{\sqrt{2}\sin\beta(\rho)}
{-e^{i \varphi}\cos\beta(\rho)}.
\end{equation}
As in the FM coreless vortex, $\beta(\rho)$ characterizes the bending
of $\nematic$
away from the $z$ axis. To form the fountain-like texture we require
$\beta(0)=\pi/2$ and increasing monotonically.
The winding number $W$ associated with the nematic texture may
be defined by taking $\nhat_F=\nematic$ in
Eq.~(\ref{eq:coreless-charge}).
Note that due to the equivalence $\nematic
\leftrightarrow -\nematic$ the sign
of $W$ is no longer well defined.
For the cylindrically symmetric fountain
texture~\eqref{eq:nematic-fountain}, the integral in
Eq.~\eqref{eq:coreless-charge} can be evaluated to yield
\begin{equation}
  \label{eq:nematic-charge}
  W = \frac{1-\cos\beta^\prime(R)}{2}
  = \frac{1-\sin\beta(R)}{2}\,,
\end{equation}
where $\beta^\prime=\beta-\pi/2$ is the angle between $\nematic$ and
the $z$ axis, such that $\beta^\prime=0$ on the symmetry axis where
$\nematic=\zhat$.
The
value of $\nematic$ is not fixed at the boundary, and the coreless
nematic texture
may smoothly dissolve. Unlike the spin texture of the FM coreless
vortex, it cannot
be stabilized by rotation, due to its vanishing mass circulation.

\subsection{Preparation of vortex states}
\label{sec:preparation}

Several methods for controlled creation of vortex states by
transferring angular momentum from an electromagnetic field have been
proposed
theoretically~\cite{marzlin_prl_1997,bolda_pla_1998,williams_nature_1999,nakahara_physicab_2000,isoshima_pra_2000,dutton_prl_2004}.
Some of these methods
have also been implemented in experiment~\cite{matthews_prl_1999,leanhardt_prl_2002,shin_prl_2004,andersen_prl_2006,pugatch_prl_2007}.
Specifically in spinor
BECs, switching of a magnetic field along the symmetry axis of the
trap has been used to create
coreless~\cite{leanhardt_prl_2003,choi_njp_2012} and
nematic coreless~\cite{choi_njp_2012,choi_prl_2012} vortices.
Two different methods have been demonstrated for controlled
preparation of nonsingular vortices. Here we give a brief
overview of each.

In Refs.~\cite{leanhardt_prl_2003,choi_njp_2012} a coreless vortex
was prepared using a time-dependent
magnetic field to induce spin rotations. This technique was first proposed
theoretically in Ref.~\cite{isoshima_pra_2000} and was also
implemented experimentally to prepare singly and
doubly quantized vortices in a spin-polarized
BEC~\cite{leanhardt_prl_2002,shin_prl_2004}.

The creation of a coreless vortex in the spin-1 BEC begins with a
condensate prepared in a fully spin-polarized state, which
we take to be $\zeta^{(1)}=(-1,0,0)^T$~\footnote{The experiment in
  Ref.~\cite{choi_njp_2012} actually starts from $\zeta=(0,0,1)^T$ and
  creates an ``upside-down'' coreless vortex.}.
The condensate is subject to an external
3D magnetic quadrupole
field~\cite{choi_njp_2012}
\begin{equation}
  \label{eq:quadrupole}
  \mathbf{B} = B^\prime \rho\rhohat + \left[B_z(t)-2B^\prime z\right]\zhat\,,
\end{equation}
where we have introduced cylindrical coordinates $(\rho,\varphi,z)$.
The zero-field point $z=B_z/2B^\prime$ ($\rho=0$) of the
quadrupole field is initially at large $z$ so that $\mathbf{B}
\parallel \zhat$ in the condensate.

The coreless-vortex structure
is created by linearly sweeping $B_z(t)$ so that the
zero-field point passes through the condensate. The changing $B_z$
causes the magnetic field away from the $z$ axis to rotate around
$\phihat$ from the $\zhat$ to the $-\zhat$ direction. The rate of
change of the magnetic field decreases with the distance $\rho$ from
the symmetry axis. Where the rate of change is sufficiently slow, the
atomic spins adiabatically follow the magnetic field,
corresponding to a complete transfer from $\zeta_+$ to $\zeta_-$ in
the laboratory frame. However, where the rate of change of the
magnetic field is rapid, atomic spin rotation is no longer adiabatic. In
the laboratory frame, the spins
thus rotate through an angle $\beta(\rho)$, given
by the local adiabaticity of the magnetic-field sweep,
which increases monotonically from zero on the symmetry
axis. Linearly ramping $B_z(t)$ thus directly implements the spin
rotation
\begin{equation}
  \label{eq:fm-cl}
  \zeta^{\rm i} =
  e^{-i\F\cdot\beta(\rho)\phihat}\zeta^{(1)} = \zeta^\mathrm{cl},
\end{equation}
where $\zeta^\mathrm{cl}$ is given by Eq.~\eqref{eq:cl}, yielding the
fountainlike spin texture~\eqref{eq:fm-fountain}
that defines the coreless
vortex in the spinor BEC.
In Ref.~\cite{choi_prl_2012} the same magnetic-field rotation
technique was applied to a condensate initially in the polar
state $\zeta=(0,1,0)^T$ to create the nematic coreless vortex,
Eq.~\eqref{eq:ncv}.

The first controlled preparation of a nonsingular
vortex~\cite{leanhardt_prl_2003} used a
2D quadrupole field together with an axial bias field.
The magnetic field in the trap is then
$\mathbf{B}(\rho,\varphi,\theta) = B_z(t)\zhat +
B^\prime\rho\left[\cos(2\varphi)\rhohat -
\sin(2\varphi)\phihat\right]$. By the mechanism described above,
ramping of $B_z(t)$ then causes a spin rotation $\zeta(\mathbf{r}) =
  \exp[-i\F\cdot\beta(\rho)\nhat]\zeta^{(1)}$ about an axis
$\nhat(\varphi)=\sin\varphi\xhat+\cos\varphi\yhat$. The rotation
yields a
nonsingular spin texture exhibiting a cross disgyration, instead of
the fountainlike structure. The two are topologically equivalent.

Another technique for phase imprinting a coreless vortex was recently
demonstrated in Ref.~\cite{leslie_prl_2009}. In this
experiment, the coreless vortex was created in the
$\ket{m=\pm2}$ and
$\ket{m=0}$ magnetic sublevels of the spin-2 manifold of
$^{87}$Rb. The phase imprinting starts with a spin-polarized condensate in
the $\ket{m=+2}$ level, with a magnetic field along the $z$ axis.
Collinear $\sigma^-$ and $\sigma^+$ polarized laser beams along the
symmetry axis then couple $\ket{m=2}$ to the $\ket{m=0}$ and
$\ket{m=-2}$ levels. The laser beams have Laguerre-Gaussian and
Gaussian intensity profiles, respectively, so that the population
transferred to the $\ket{m=0}$ ($\ket{m=-2}$) level picks up a $2\pi$
($4\pi$) phase winding. The intensity minimum of the Laguerre-Gaussian
beam leaves a 
remaining population in $\ket{m=2}$ with no phase winding. The
resulting five-component spinor represents a coreless vortex with
the spin structure~\eqref{eq:fm-fountain}
when the three nonempty
levels of the five-component spinor are regarded as a (pseudo)spin-1
system.
The bending angle $\beta$ is determined by the density
profiles of the nonempty spinor components. The laser beams inducing
the Raman coupling of the magnetic sublevels can be tailored with a
high degree of control, and the vortex structure can therefore be
precisely engineered.

By accurately creating specific spin textures, phase
imprinting of coreless vortices
gives control over the longitudinal magnetization
of the cloud, regardless of whether interactions are polar or FM.
In the spin-2 coreless-vortex experiment~\cite{leslie_prl_2009},
the resulting magnetization in the spin-2 manifold is measured at
$M=0.64$ for an imprinted Anderson-Toulouse-Chechetkin-like spin
texture, and at $M=0.72$ 
for a Mermin-Ho-like texture.
In the magnetic-field rotation experiment~\cite{leanhardt_prl_2003}
the local magnetization
$\M(\rr)=[n_+(\rr)-n_-(\rr)]/n(\rr)$ is reported to be
$\sim 0.7$ at the center of the cloud and
$\sim -0.5$
at the edge. Because of the lower density in the negatively
magnetized region, also this vortex can be estimated to carry a
positive, nonzero magnetization $M$.

Note that the spinor vortices presented above are composed of vortex
lines in the individual components of the spinor order parameter.
By phase-imprinting these vortex lines using existing
techniques~\cite{matthews_prl_1999,leanhardt_prl_2002,andersen_prl_2006}
it would be possible to prepare also singular vortex states.

\section{Vortex-Core Wave Functions}
\label{sec:interpolation}

The two phases of the spin-1 BEC have different order-parameter
symmetries that support different topological defects. For
an overview, see, e.g.,
Refs.~\cite{kawaguchi_physrep_2012,borgh_pra_2013}. Here we are
interested in vortex states that mix the
two phases. Such solutions representing coreless and nematic-coreless
vortex textures appearing in the cores of singly quantized vortices
were presented in Ref.~\cite{lovegrove_prl_2014}.  Here we expand the
discussion and provide the full derivation and additional examples of
phase-mixing vortex wave functions.
For that purpose we write the spinor wave function as
\begin{equation}
  \label{eq:general}
  \zeta = \frac{e^{i\tau}}{2}
    \threevec{\sqrt{2}e^{-i\alpha} \left( e^{i\gamma}
      D_-\sin^2\frac{\beta}{2}
      -e^{-i \gamma}D_+\cos^2\frac{\beta}{2}\right)}
     {-\left(e^{i\gamma}D_- + e^{-i \gamma}D_+\right)\sin\beta}
     {\sqrt{2}e^{i \alpha}\left(e^{i\gamma}D_-\cos^2\frac{\beta}{2}
       - e^{-i \gamma}D_+\sin^2\frac{\beta}{2}\right)},
\end{equation}
where $D_\pm=\sqrt{1 \pm F}$ is given in terms of the spin magnitude $\absF=F$.
Here $\tau,\alpha,\beta,\gamma$ are (in general spatially varying)
parameters that can be used to determine the specific vortex states
that we want to analyze. In the vortex states that smoothly
interpolate between the FM and polar phases also $F$ will vary in
space. The state~\eqref{eq:general} can 
also be specified by the condensate phase, the spin magnitude $F$ and an
orthonormal triad with one vector in the direction of the spin.
One of the remaining vectors in the triad forms the nematic axis
$\nematic$. (In the polar limit,
$\nematic$ fully specifies the
state together with the condensate
phase~\cite{ruostekoski_prl_2003}.) 

In order to understand the origin of Eq.~\eqref{eq:general}, one
should consider stationary, uniform, vortex-free spinor wave functions
in the presence of Zeeman energy
shifts~\cite{zhang_njp_2003,murata_pra_2007,ruostekoski_pra_2007}. For
instance, for the uniform spin profile $\inleva{\F}=F\zhat$ we take 
\begin{equation}
  \label{eq:zetaF}
  \zeta^F=\frac{1}{\sqrt{2}}\threevec{-\sqrt{1+F}}{0}{\sqrt{1-F}}.
\end{equation}
In this case $\nematic = \xhat$. The limits
$F=0$ and $F=1$ yield the polar and FM phases,
$\zeta^F|_{F=0}=(-1/\sqrt{2},0,1/\sqrt{2})^T$ and
$\zeta^F|_{F=1}=(-1,0,0)^T$, respectively.
The general spinor with $\absF=F$ can then be reached by applying a
condensate phase $\tau$ and a 3D spin rotation
$U(\alpha,\beta,\gamma) =
\exp(-i\F_z\alpha)\exp(-i\F_y\beta)\exp(-i\F_z\gamma)$, defined
by three Euler angles, resulting in the spatially uniform version of
Eq.~\eqref{eq:general}.

\subsection{Vortex core filling}
\label{sec:vortex-core-filling}

The FM and polar phases of the spin-1 BEC mix when the energy of
singular vortices relaxes. As discussed in section~\ref{sec:mft}, when
the spin healing length exceeds the density healing length, it is
energetically favorable for the vortex core to fill with atoms in the
opposite phase. We can now explicitly construct vortex wave functions
that represent the full, continuous interpolation between the vortex
and its filled, nonrotating core.

\subsubsection{Polar vortices}

Consider a singly quantized vortex in a polar condensate.
When the core fills, the wave function must
reach the FM phase on the singularity of the polar order parameter.
Consider the choice  $\tau=\gamma=\varphi$ and $\alpha=0$, with
constant $\beta=\beta_0$, in Eq.~\eqref{eq:general}. We then obtain
\begin{equation}
  \label{eq:101-general}
  \zeta=\frac{e^{i\varphi}}{2}
    \threevec{\sqrt{2} \left(e^{ i \varphi} \sin^2\frac{\beta_0}{2}D_-
              -e^{-i \varphi}\cos^2\frac{\beta_0}{2}D_+\right)}
	     {\sin\beta_0\left(e^{i \varphi} D_-+e^{-i \varphi} D_+\right)}
	     {\sqrt{2}\left(e^{i  \varphi} \cos^2\frac{\beta_0}{2}D_-
	      - e^{-i \varphi}\sin^2\frac{\beta_0}{2}D_+\right)},
\end{equation}
which reduces to Eq.~\eqref{eq:p012}, with $\gamma_0=0$,
in the $F=0$ limit, representing a singly quantized polar vortex
with a $2\pi$ winding in $\nematic$. In the limit $F=1$, on the
other hand, Eq.~\eqref{eq:101-general} represents the vortex-free
FM phase. By allowing $F(\rho)$ to decrease monotonically
from $F(\rho\to0)=1$ to $F=0$ away from the vortex line, we find the
spinor representing a polar vortex with a FM core.

We can confirm this interpretation by studying the superfluid
circulation. A general expression for the superfluid velocity may be
derived from Eq.~(\ref{eq:general}) as
\begin{equation}
  \label{eq:sf-velocity}
  \mathbf{v} =
    \frac{\hbar}{m\rho}
    \left(\grad\tau-F\grad\gamma-F\cos\beta\grad\alpha\right).
\end{equation}
Since we here assume axial symmetry, we consider the mass circulation
on a path at constant $\rho$,
\begin{equation}
  \label{eq:nu-winding}
  \nu =
  \oint d\mathbf{r} \cdot \mathbf{v} =
  \frac{h}{2m}\left(l-qF-pF\cos\beta\right),
\end{equation}
where in our construction $l=2d\tau/d\varphi$, $p=2d\gamma/d\varphi$,
and $q=2d\alpha/d\varphi$ are integers.
For the vortex~\eqref{eq:101-general}, we then obtain
$\nu=h(1-F)/m$, and we see that this interpolates smoothly between the
non-circulating FM core ($F=1$) and the single quantum of circulation
in the polar phase ($F=0$).  Both numerical
simulations~\cite{lovegrove_pra_2012} and experimental
observations~\cite{seo_prl_2015} 
have demonstrated how the axisymmetry of this singly quantized vortex
breaks under dissipation and leads to the formation of a pair of
half-quantum vortices.

The half-quantum vortex, described by Eq.~\eqref{eq:hq}, is likewise
singular, and its core may fill with the FM phase.
This state may be constructed from Eq.~\eqref{eq:general} by choosing
$\tau=\varphi/2$, $\gamma=\pi+\varphi/2$
and $\alpha=0$ with constant $\beta=\beta_0$ to yield
\begin{equation}
  \label{eq:hq-general}
  \zeta=\frac{e^{i\frac{\varphi}{2}}}{2}
      \threevec{\sqrt{2}\left(e^{-i\frac{\varphi}{2}}\cos^2\frac{\beta_0}{2}D_+
	         - e^{i\frac{\varphi}{2}}\sin^2\frac{\beta_0}{2}D_-\right)}
	       {\sin\beta_0
		 \left(e^{-i\frac{\varphi}{2}}D_+
		 +e^{i\frac{\varphi}{2}}D_-\right)}
	       {\sqrt{2}\left(e^{-i\frac{\varphi}{2}}\sin^2\frac{\beta_0}{2}D_+
		 -e^{i\frac{\varphi}{2}}\cos^2\frac{\beta_0}{2}D_-\right)}.
\end{equation}
The $F=1$ limit is the vortex-free FM phase with
$\expF=\sin\beta_0\xhat+\cos\beta_0\zhat$. By allowing $F(\rho)$ to decrease
monotonically with the radial distance from $F=1$ to $F=0$, the
circulation, $\nu=h(1-F)/2m$ from Eq.~\eqref{eq:nu-winding},
smoothly interpolates between the inner, non-circulating FM phase and the
outer polar half-quantum vortex.

\subsubsection{FM vortices}

Also in the FM phase, cores of singular vortices may be filled (with
atoms in the polar phase) through the same mechanism.
In Eq.~\eqref{eq:spinvortex}, representing a
singular FM vortex, $\zeta_0$ is
nonsingular everywhere, and can therefore fill the vortex core.
We now generalize this
solution to a spinor that interpolates between an outer singular FM vortex
and an inner polar core. We choose $\alpha=\varphi$ and $\tau=0$
with arbitrary
$\gamma=\gamma_0$, yielding
\begin{equation}
  \label{eq:spinvortex-general}
  \zeta= \frac{1}{2}
     \threevec{\sqrt{2}e^{-i \varphi}
                \left(e^{ i \gamma_0} \sin^2\frac{\beta}{2}D_-
		-e^{-i \gamma_0}\cos^2\frac{\beta}{2}D_+\right)}
	      {-\sin\beta\left(e^{i \gamma_0} D_-
		+e^{-i \gamma_0} D_+  \right)}
	      {\sqrt{2}e^{i \varphi}
		\left(e^{i\gamma_0} \cos^2\frac{\beta}{2}D_-
		- e^{-i \gamma_0}\sin^2\frac{\beta}{2}D_+\right)}.
\end{equation}
In the $F=0$ limit, the corresponding circulation $\nu= -h F\cos\beta/m$
vanishes, and the spinor represents a non-circulating polar condensate
with $\nematic=\cos\beta\rhohat-\sin\beta\zhat$.

\subsection{Composite topological defects}
\label{sec:composite-defects}

\begin{table*}[tb]
\begin{tabular}{l|l|l|l|l|l}
FM limit & Polar limit & $\tau/\varphi$ & $\alpha/\varphi$ & $\gamma/\varphi$ & $m\nu/h$\\
\hline
No Vortex & Half-Quantum Vortex & $1/2$ & $0$ & $1/2$ & $(1-F)/2$\\
No Vortex & Singly Quantized Vortex & $1$ & $0$ & $1$ & $1-F$\\
Coreless Vortex & Half-Quantum Vortex & $1/2$ & $1$ & $-1/2$ & $(1+F)/2-F\cos\beta$\\
Coreless Vortex & Singly Quantized Vortex & $1$ & $1$ & $0$ & $1-F\cos\beta$\\
Coreless Vortex & Nematic Coreless Vortex & $0$ & $1$ & $-1$ & $F(1-\cos\beta)$\\
Singular Vortex & No Vortex & $0$ & $1$ & $0$ & $-F\cos\beta_0$\\
Singular Vortex & Nematic Coreless Vortex & $0$ & $1$ & $0$ & $-F\cos\beta(\rho)$\\
Singular Vortex & Half-Quantum Vortex & $1/2$ & $1$ & $1/2$ & $(1-F)/2-F\cos\beta$\\
Singular Vortex & Half-Quantum Vortex & $1/2$ & $0$ & $-1/2$ & $(1+F)/2$\\
Singular Vortex & Singly Quantized Vortex & $1$ & $0$ & $0$ & $1$
\end{tabular}
\caption{Spinor wave functions for vortex states that continuously
  connect FM and 
  polar phases can be constructed from
  Eq.~(\ref{eq:general}). The table shows the state in the FM and
  polar limits and the choices for $\tau$, $\alpha$ and $\gamma$
  necessary to construct them, given as half-integer or integer
  multiples of the azimuthal angle $\varphi$. The right-most column
  shows the corresponding superfluid circulation $\nu$ as a function of spin
  magnitude $F$ and the second Euler angle $\beta$ (also indicating
  any assumed functional form for nonconstant $\beta$ where
  required). Note that vortex-free FM (polar) limits provide wave
  functions that simultaneously describe a singular polar (FM) vortex
  and its filled core, while solutions that exhibit vortex states in
  both limits correspond to composite defects.}
\label{tab:analytics}
\end{table*}

The vortex wave functions constructed above, simultaneously representing a
singular vortex and its core, are summarized in the first, second and
sixth rows of Table~\ref{tab:analytics}.  We give the corresponding
choices for the condensate phase and spin rotations as well as the
mass circulation.

We now generalize the construction of phase-mixing spinors to allow
vortex states in both polar and FM limits.  The wave function may
then, for example, represent a defect configuration that exhibits different
small- and large-distance topology of the vortex core. Such \emph{composite
  topological defects} exhibit a hierarchy of core
structures, between which the wave function interpolates smoothly with
radial distance from the vortex line. We provide explicit examples with
both singular and nonsingular inner cores as shown in Table~\ref{tab:analytics}.
The interpolation between the two solutions is reminiscent of the interfaces between spatially separated, coexisting phases of a
superfluid system. Constructions analogous to those in
Table~\ref{tab:analytics} can then be used to describe wave functions
that continuously connect defects across the topological
interface~\cite{borgh_prl_2012,borgh_njp_2014}.

By making appropriate choices for the Euler angles and condensate
phase in Eq.~\eqref{eq:general}, a large family of nontrivial wave
functions connecting polar and FM vortex states may thus be
constructed.  In Table~\ref{tab:analytics} we list the basic FM- and
polar-limit vortex states and provide the $\tau$, $\alpha$, and
$\gamma$
needed to construct them from
Eq.~\eqref{eq:general} as half-integer (or integer,
for $\alpha$) multiples of the azimuthal angle $\varphi$. The
right-most column gives the superfluid circulation as a function of
$F$ and $\beta$.  Note that in addition to the filled-core vortex
states, we are now able construct inner cores with nontrivial,
nonsingular textures. Such states become relevant for example when a
FM coreless vortex is phase-imprinted on a polar
condensate~\cite{lovegrove_prl_2014}. In addition, we also find more
exotic wave functions that connect singular vortices, as well as a
smooth connection of nontrivial, nonsingular textures.

In sections~\ref{sec:coreless} we will
demonstrate numerically how, in addition to the filled cores of
singular vortices (section~\ref{sec:weak-mag}), a singly quantized
polar vortex with
coreless-vortex FM core may be energetically stable, and that a FM
vortex with nematic-coreless-vortex core may be stable in an effective
two-component 
regime at strong magnetization.
Singular
vortices with singular inner cores do not naturally appear as the
ground state, but we will show how they may be stabilized through
precise control over the quadratic Zeeman shift in
section~\ref{sec:composite-core}.

In the following we will analyze in detail the construction of those
composite defect states particularly relevant for the analysis of our
numerical results.  Some additional constructions and discussion are
given in the Appendix.

\subsubsection{Singly quantized polar vortex with coreless-vortex core}

As a first example, we construct a spinor wave function representing a
FM coreless vortex that appears as the core of a polar vortex, as
given in Ref.~\cite{lovegrove_prl_2014}.  This
vortex structure is analyzed numerically in
section~\ref{sec:coreless-polar-regime}.
In Eq.~(\ref{eq:general}) we choose $\tau=\alpha=\varphi$ and
$\gamma=0$ to yield:
\begin{equation}
  \label{eq:cl-p012}
    \zeta^{\rm cl} =
    \frac{1}{2}
    \threevec{\sqrt{2}\left(D_-\sin^2\frac{\beta}{2}
      - D_+\cos^2\frac{\beta}{2}\right)}
             {-e^{i\varphi}(D_- + D_+)\sin\beta}
             {\sqrt{2}e^{2i\varphi}\left(D_-\cos^2\frac{\beta}{2}
	       -D_+\sin^2\frac{\beta}{2}\right)}.
\end{equation}
The spin texture is then given by
\begin{equation}
  \label{eq:spin-texture}
  \inleva{\F} =
     F(\rr)[\sin\beta(\rr)\rhohat + \cos\beta(\rr)\zhat]\,,
\end{equation}
where $\beta(\rr)$ increases monotonically from zero on the
symmetry axis to form the characteristic fountain texture with
  varying spin magnitude $F(\rr)$.
In the limit $F=1$, we retrieve the coreless vortex
represented by Eqs.~\eqref{eq:cl} and \eqref{eq:fm-fountain}.
In the polar limit $F=0$, on the other hand, Eq.~\eqref{eq:cl-p012}
represents a singly quantized vortex
\begin{equation}
  \label{eq:polar-singular}
  \left.\zeta^{\rm cl}\right|_{F\to0} =
    \frac{e^{i\varphi}}{\sqrt{2}}
    \threevec{-e^{-i\varphi}\cos\beta}
             {-\sqrt{2}\sin\beta}
             {e^{i\varphi}\cos\beta}\,,
\end{equation}
where we have explicitly separated out the condensate phase
$\tau=\varphi$. The nematic axis forms the texture
$\nematic = \cos\beta\rhohat - \sin\beta\zhat$. In general the spin
rotation that accompanies the winding of the condensate phase
therefore represents a disgyration of the nematic axis.

The vortex \eqref{eq:cl-p012} can represent a solution for which $F$ is
nonuniform, so that
Eqs.~\eqref{eq:cl} and~\eqref{eq:polar-singular} are the two
limiting solutions. We can form a composite topological defect by
setting $F(\rho=0)=1$ and $\beta(\rho=0)=0$ at the center and letting
$F\rightarrow0$ and $\beta\rightarrow \pi/2$ as $\rho$ increases. Then
the core exhibits a coreless-vortex fountain texture that
continuously transforms toward a singular polar vortex as the radius
increases.

The mixing of the polar and FM phases in the vortex configuration is
also reflected in the superfluid velocity given by
Eq.~\eqref{eq:sf-velocity}.
For the vortex state~\eqref{eq:cl-p012}, this reduces to
\begin{equation}
  \label{eq:cl-v}
  \mathbf{v}^{\rm cl} =
    \frac{\hbar}{m \rho}\left[1 - F(\rho)\cos \beta(\rho)\right]\phihat\,,
\end{equation}
when $F$ and $\beta$ depend only on the radial distance $\rho$.
By considering a circular loop $\mathcal{C}$ at constant $\rho$
enclosing the vortex line,
we can then compute the circulation
\begin{equation}
  \label{eq:circ}
  \nu = \int_\mathcal{C} d\mathbf{r} \cdot \mathbf{v}^{\rm cl}
  =\frac{h}{m}\left[1 - F(\rho)\cos\beta(\rho)\right].
\end{equation}
Note that for nonzero $F$, circulation increases with increasing
$\beta(\rho)$, implying that the coreless-vortex texture can adapt
to an imposed rotation. This indicates that the spin texture will
bend more sharply at faster rotation to provide increased angular
momentum. We may regard the integrand of Eq.~\eqref{eq:circ} as a
\emph{circulation density}
\begin{equation}
  \label{eq:circ-dens}
  \mathcal{V}(\rr) = \mathbf{v}(\rr)\cdot\phihat\rho\,
\end{equation}
along a cylindrically symmetric path. The circulation of Eq.~(\ref{eq:cl-p012})
continuously interpolates between the polar and FM phases, smoothly
connecting the small-distance
and large-distance topology of the vortex. Note
that it further follows from Eq.~\eqref{eq:circ} that circulation alone
is quantized only in the limit $F\rightarrow 0$.

Note that the winding number $W$, given by
Eq.~\eqref{eq:coreless-charge}, remains defined also for the spin
texture of the composite defect, provided that the surface
$\mathcal{S}$ covers the full cross section of the non-polar core.
By substituting $\inleva{\F}$ from Eq.~(\ref{eq:spin-texture})
into Eq.~(\ref{eq:coreless-charge}), we may evaluate $W$. Assuming
cylindrical symmetry and taking $R$ to be the radial extent of the
spin texture, $W$ is again given by Eq.~\eqref{eq:cl-W}.

\subsubsection{Singular FM vortex with nematic-coreless-vortex core}

In section~\ref{sec:ncv} we numerically investigate the stability
of the nematic coreless vortex constructed in
Eq.~\eqref{eq:ncv}. 
In order for the vortex to be stabilized by
rotation, the condensate must develop nonpolar, circulation-carrying
regions.  We therefore generalize the
nematic coreless vortex solution to the spinor wave function given in
Ref.~\cite{lovegrove_prl_2014} that also allows nonzero spin.
We note that
we wish to construct a vortex where
\begin{equation}
  \label{eq:nematic-fountain}
  \nematic = \sin\beta^\prime(\rho)\rhohat + \cos\beta^\prime(\rho)\zhat\,,
\end{equation}
corresponding to the state phase-imprinted by Choi et
al.~\cite{choi_prl_2012,choi_njp_2012}. The angle $\beta^\prime$
between $\nematic$ and the $z$ axis increases from $\beta^\prime=0$ at
$\rho=0$ to $\beta^\prime=\pi/2$ ($\beta^\prime=\pi$) at the edge
for a Mermin-Ho-like (Anderson-Toulouse-Chechetkin-like) texture.
Note that since the Euler angles in Eq.~(\ref{eq:general}) represent
spin rotations of Eq.~\eqref{eq:zetaF}, we have
$\beta(\rho)=\beta^\prime(\rho)+\pi/2$, such that $\beta = \pi/2$ at
the center of
the vortex. The desired vortex state can then be constructed from
Eq.~\eqref{eq:general} by
additionally choosing $\alpha=\varphi$, $\gamma=\pi$ and
$\tau=0$ to yield
\begin{equation}
  \label{eq:nematic-vortex}
  \zeta^{\rm n}= \frac{1}{2}
    \threevec{\sqrt{2}e^{-i\varphi}\left(D_+\cos^2\frac{\beta}{2}
               -D_-\sin^2\frac{\beta}{2}\right)}
	     {\left(D_+ + D_-\right)\sin\beta}
	     {\sqrt{2}e^{i\varphi}\left(D_+\sin^2\frac{\beta}{2}
	       -D_-\cos^2\frac{\beta}{2}\right)},
\end{equation}
with spin profile
$\inleva{\F}=F(\sin\beta(\rho)\rhohat+\cos\beta(\rho)\zhat)$. Note
that this is a special case of Eq.~\eqref{eq:spinvortex-general}, with
the angle $\beta(\rho)$ chosen as a monotonically increasing function
of $\rho$ to yield the desired fountain
texture~\eqref{eq:nematic-fountain} in $\nematic$.

In a magnetized BEC, Eq.~\eqref{eq:nematic-vortex} can represent a composite
vortex that mixes the FM and polar phases. We consider a solution for which
$F$ exhibits a spatial structure interpolating between $F\rightarrow0$
at the center and $F\rightarrow 1$
at the edge of the cloud. In the limit $F\rightarrow 1$,
Eq.~\eqref{eq:nematic-vortex} becomes
a singular singly quantized FM vortex,
\begin{equation}
  \left.\zeta^{\rm n}\right|_{F\to1}= \frac{1}{\sqrt{2}}
    \threevec{\sqrt{2}e^{-i\varphi}\cos^2\frac{\beta}{2}}
	     {\sin\beta}
	     {\sqrt{2}e^{i\varphi}\sin^2\frac{\beta}{2}}.
\end{equation}

\subsubsection{Singular FM vortex with singular polar-vortex core}

It is also possible to construct composite topological defects where
both the inner and outer vortices are singular (leaving open the
question of the core structure of the inner singular vortex). We
demonstrate in section~\ref{sec:composite-core} how such structures
may be stabilized using a tunable quadratic Zeeman shift.
The simplest example to construct is that of a singular FM vortex
[Eq.~\eqref{eq:fmsingular}], forming the
core of the singly quantized polar vortex [Eq.~\eqref{eq:psingular}], or
vice versa. (The more complicated connection of a singular FM vortex
to a polar half-quantum vortex is provided in
Appendix~\ref{app:further-examples}.)
We construct the corresponding spinor by
setting $\tau=\varphi$ and $\alpha=0$. For any constant
$\gamma=\gamma_0$ and $\beta=\beta_0$, we then
have
\begin{equation}
  \label{eq:phasevortex-general}
  \zeta= \frac{e^{i\varphi}}{2}
    \threevec{\sqrt{2}\left(e^{i\gamma_0}\sin^2\frac{\beta_0}{2}D_-
               -e^{-i \gamma_0}\cos^2\frac{\beta_0}{2}D_+\right)}
	     {-\sin\beta_0\left(e^{i \gamma_0} D_-
	       +e^{-i \gamma_0} D_+  \right)}
	     {\sqrt{2}\left(e^{i\gamma_0}\cos^2\frac{\beta_0}{2}D_-
	       -e^{-i \gamma_0}\sin^2\frac{\beta_0}{2}D_+\right)},
\end{equation}
with circulation $\nu=h/m$. This vortex is singular for all values of
$F$ and $\beta$, such that the singularity cannot be avoided by
judicious choice of parameters. In the $F=1$ limit then,
Eq.~\eqref{eq:phasevortex-general} describes a singular FM vortex,
with uniform spin texture
\mbox{$\expF=\sin\beta_0 \xhat + \cos\beta_0\zhat$}.
The $F=0$ limit similarly corresponds to a singular polar vortex
with uniform
$\nematic=\cos\beta_0\cos\gamma_0\xhat+\sin\gamma_0\yhat-\sin\beta_0\cos\gamma_0\zhat$.
A wave function that continuously interpolates between the two singular
vortices is then constructed by taking $F(\rho)$ to vary monotonically
between $F=0$ and $F=1$. The boundary condition on $F(\rho)$ away
from the vortex line determines the large-distance topology.

\section{Nonsingular vortices and textures}
\label{sec:coreless}

Nonsingular vortices have been prepared in several recent
experiments~\cite{leanhardt_prl_2003,leslie_prl_2009,choi_prl_2012,choi_njp_2012}. 
However, in Refs.~\cite{leanhardt_prl_2003,choi_njp_2012} a FM
coreless vortex was prepared (using the magnetic-field rotation
technique described in section~\ref{sec:preparation}) in $^{23}$Na,
which exhibits polar interactions, in which 
case the vortex would not be expected to be stable by simple energetic
arguments alone.  
In Ref.~\cite{lovegrove_prl_2014}, we demonstrated that conservation
of magnetization can stabilize the imprinted coreless vortex, and at
large magnetization also a nematic coreless vortex. We noted that it
may also destabilize the coreless vortex in a FM condensate.
In the following, we first expand the discussion of the coreless
vortex in the FM interaction regime and determine in detail how the
energetically stable 
structure varies as magnetization is conserved at different values. In
particular, we explain how conservation of a sufficiently weak
magnetization can destabilize the coreless vortex texture, such that a
singular vortex instead forms the ground state in the rotating FM condensate.
We then
explain how conservation of magnetization can also energetically stabilize the
coreless vortex when it is phase-imprinted on a BEC in the polar
regime~\cite{lovegrove_prl_2014}.
Equation~\eqref{eq:p012} will then allow us to understand the stable
coreless vortex as the composite core of a singly quantized vortex in
the polar BEC. 
We also investigate whether conservation of
magnetization can stabilize a nematic coreless vortex, and explain our
finding~\cite{lovegrove_prl_2014}  that stability is achieved only at
strong magnetization in an effective two-component regime.

\subsection{Coreless Vortex}
\label{sec:coreless-vortex}

\subsubsection{FM Regime}
\label{sec:coreless-fm-regime}

We first study the stability and spin texture of the coreless vortex
as it relaxes under conservation of magnetization
in the FM interaction regime.
The initial state in our numerical energy relaxation is the coreless vortex
as created in phase-imprinting experiments, Eq.~\eqref{eq:cl}.
The initial state is in the FM phase everywhere,
with $\beta(\rho)$ specifying the initial magnetization. In the FM interaction
regime, the condensate remains in the  FM phase with $F=1$ everywhere
during relaxation when magnetization is not conserved. We 
find however that conservation of a sufficiently weak magnetization
can lead to energetic 
instability of the coreless vortex, with a singular vortex becoming the
rotating ground state. This may seem surprising, since in studies that
do not conserve magnetization in energy relaxation, coreless vortices
are predicted to form the ground state at sufficiently rapid
rotation~\cite{mizushima_prl_2002,mueller_pra_2004,martikainen_pra_2002,reijnders_pra_2004};
a singular vortex may then also be energetically (meta)stable,
but always has a higher energy than the coreless
vortex~\cite{lovegrove_pra_2012}.  To understand
this instability of the coreless vortex, it is instructive to first consider
the stable configuration of a FM coreless vortex when magnetization is not
explicitly conserved in the relaxation process.
\begin{figure}[tb]
  \centering
  \includegraphics[width=0.92\columnwidth]{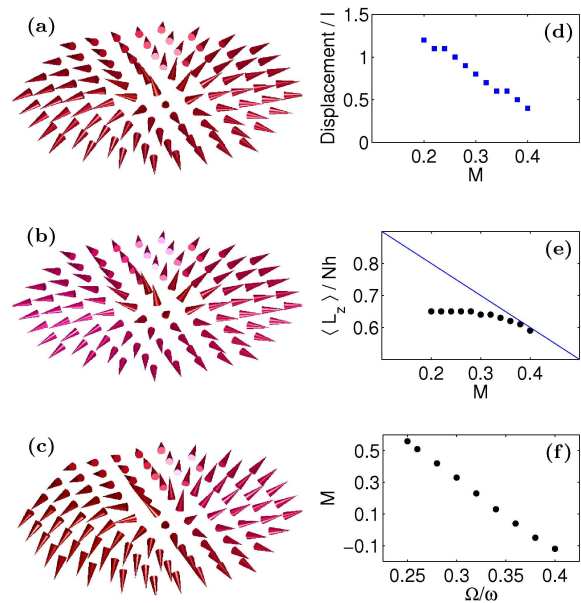}
  \caption{Numerically
    calculated spin textures in the $x$-$y$
    plane for the coreless vortex in the FM regime with an initial magnetization
    $M=0.4$ (a) not conserved and (b) conserved throughout the
    relaxation process. (c) The same for a conserved magnetization of $M=0.2$,
    showing a displacement of the coreless vortex relative to the more
    strongly-magnetized
    case. (d) Displacement and (e) angular momentum of the coreless
    vortex in a trap
    rotating at $\Omega/\omega=0.3$ for
    different values of the conserved magnetization (black dots) compared
    with the angular momentum of an axisymmetric coreless vortex at
    the center of the trap,
    for the same magnetization (blue line). (f) Numerically calculated
    magnetization of the energy-minimizing
    coreless vortex in the FM regime as a function of rotation frequency,
    where magnetization-conserving relaxation has not been enforced.
  }
  \label{fig:fm-coreless-weak}
\end{figure}

When the magnetization is not conserved, the fountain texture
of the FM coreless vortex displays a characteristic radial profile
of $\beta(\rho)$. As the rotation frequency increases, the angular
momentum also increases. Since increased angular momentum
requires a sharper bending of $\beta(\rho)$, the result is that
the magnetization of the energetically stable configuration decreases
as the rotation frequency increases, as illustrated in
Fig.~\ref{fig:fm-coreless-weak}(f).

To understand this, we present a qualitative description of an axisymmetric
vortex at the center of the trap, described by the spinor,
\begin{equation}
    \zeta= e^{i a \varphi}
    \threevec{e^{-ib \varphi}\abs{\zeta_+(\rho)}}
             {\abs{\zeta_0(\rho)}}
             {e^{ib \varphi}\abs{\zeta_-(\rho)}}\,,
\end{equation}
where $\inlabs{\zeta_i(\rho)}^2$ are the populations of the three
spinor components as a function of radius, 
giving rise to the radial profile of $\beta$. The integers $a$ and
$b$ represent the winding of the condensate 
phase and spin vector respectively. The expectation value of the
angular momentum for such a vortex is~\cite{pethick-smith},
\begin{equation}
\inleva{\hat{L}_z} = (a-b) N_+ + a N_0 + (a+b) N_-\,,
\end{equation}
which may be simplified via Eq.~(\ref{eq:M}) to,
\begin{equation}
\inleva{\hat{L}_z} = (a - b M)N \,.
\label{eq:genericLz}
\end{equation}
In the case of the coreless vortex, $a=b=1$ and so the angular
momentum increases linearly
with decreasing magnetization. This illustrates why increasing the
rotation frequency decreases
the magnetization when the magnetization is not conserved---increased
angular momentum acts to
decrease the longitudinal magnetization in the system. This
qualitative description is also useful
in understanding the behavior of the coreless vortex subject to
magnetization-conserving relaxation.

The fountain texture of the coreless vortex energetically favors
nonzero magnetization.
It is therefore instructive to first consider a conserved
magnetization close to that which would arise if magnetization were
unconstrained. At a trap-rotation frequency just above that at which
the coreless vortex becomes stable, $M\sim0.5$.  Strictly conserving
$M = 0.5$ throughout energy relaxation then has little impact on the
structure of the coreless vortex, as one would intuitively
expect. Keeping the rotation frequency constant, we now study the
consequences as $M$ deviates from $0.5$.

Reducing the value of the conserved magnetization leads to a
displacement of the coreless
vortex from the trap center, as demonstrated in
Fig.~\ref{fig:fm-coreless-weak}(d). This
may be understood as reducing the contribution to the magnetization
arising from the center of the
vortex, by forcing the center of the vortex to lie in a region of
lower density. At
the same time, the continuous bending of the spin vector ensures that
an enlarged region
of negative magnetization density forms at the edge of the trap
farthest from the center
of the vortex. The combination of these two effects results in a
reduction of the total
longitudinal magnetization.

\begin{figure}[tb]
  \begin{center}
    \includegraphics[width=\columnwidth]{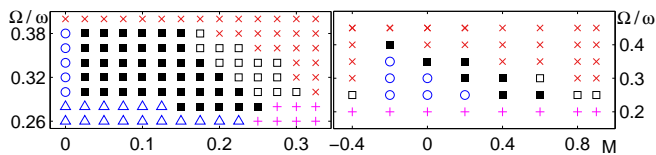}
  \end{center}
  \caption{Energetic stability of the coreless vortex in the polar (left) and
    FM (right)  interaction regimes;
    ($\blacksquare$) stable coreless vortex;
    ($\square$) stable effective two-component coreless vortex;
    ($\triangle$) instability towards a half-quantum vortex,
    ($\circ$) pair of half-quantum
    vortices (polar regime) or singular vortex (FM regime), ($+$)
    vortex-free state;
    ($\times$) nucleation of additional vortices;
  }
  \label{fig:stability}
\end{figure}

Decreasing the magnetization further causes a greater displacement
of the coreless vortex as illustrated in
Fig.~\ref{fig:fm-coreless-weak}(a-d) until,
at $M\sim0.2$, the coreless vortex is unstable towards splitting
into a pair of singular vortices. We note that the magnetization at
which this happens
decreases as the rotation frequency increases and thus infer that it
is the displacement of
the coreless vortex that triggers this instability. Further
displacement of the vortex would
produce a vortex-free state, which for a range of rotation frequencies
is higher in energy
than the singular vortex. The coreless vortex therefore splits into a
pair of singular vortices,
one of which then exits the cloud. Contrary to the findings when magnetization
is not conserved, which showed that the coreless vortex is always the
ground state, we
find that the singular vortex is in fact the ground state for
sufficiently weak magnetization. The
range of magnetization for which this is true, decreases with
increasing rotation frequency of the
trap.  The corresponding stability diagram is shown in
Fig.~\ref{fig:stability}.

One additional consequence of this displacement of the coreless vortex
due to a reduction
of the conserved magnetization, is that the angular momentum is
approximately independent
of the magnetization provided that the coreless vortex remains stable, shown in
Fig.~\ref{fig:fm-coreless-weak}(e). From the discussion
of the axisymmetric coreless vortex at the center of the trap, we saw
that decreasing the
magnetization served to increase the angular momentum. However, the
displacement of the
vortex then reduces the angular momentum relative to the axisymmetric
vortex at the center
of the trap, canceling out the increased angular momentum due to the
reduction in magnetization.

\begin{figure}[tb]
  \begin{center}
    \includegraphics[width=\columnwidth]{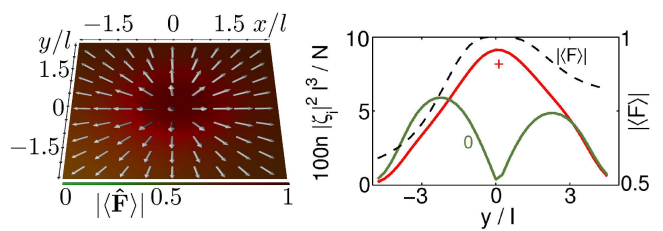}
  \end{center}
  \caption{Coreless vortex in strongly magnetized
    condensate. (Left)
    Spin magnitude (color scale) and spin vector (arrows).  Note how
    the spin magnitude falls below $1$ away from the vortex center.
    (Right) Spinor wave function along $y$-axis, showing how the strong
    magnetization leads to an effective two-component regime where
    $\zeta_-$ is depleted, and $\zeta_0$ exhibits a $2\pi$ phase
    winding.
  }
  \label{fig:two-comp}
\end{figure}

The FM coreless vortex was found to be stable for magnetization below
$M\sim0.5$.
For stronger magnetization we obtain an effective two-component
coreless-vortex state shown in Fig.~\ref{fig:two-comp},
where $\zeta_0$ represents a singly quantized vortex
whose core is filled by $\zeta_+$.
The transition to the two-component system occurs when the
$\beta(\rho)$ profile no longer
allows the three-component vortex~\eqref{eq:cl} to satisfy the
magnetization constraint, resulting in depopulated $\zeta_-$.
The threshold magnetization value decreases with rotation until
at $\Omega \simeq 0.35\omega$ the coreless vortex is stable only in the
two-component regime.  Despite the resulting formation of non-FM
regions away from the vortex line, the state nevertheless represents a
coreless vortex with a charge $W$ that may be calculated from
Eq.~\eqref{eq:coreless-charge}.

As noted above, the FM coreless vortex in the FM regime can
be unstable against
splitting into a pair of singular vortices when conservation of a weak
magnetization is imposed. One
of these vortices then exits the cloud leaving a single, singular
vortex where relaxation
without explicitly conserving magnetization would predict the coreless
vortex to be stable. As the magnetization becomes increasingly negative, one
might naturally expect also this singular vortex to become unstable as
$\zeta_+$ is depopulated. One might further
expect that the resulting vortex structure would be a singular
vortex in an effective two-component regime, exhibiting a polar vortex
core in an otherwise FM condensate.
Perhaps surprisingly, we find that a strongly magnetized coreless
vortex, similar to Fig.~\ref{fig:two-comp} but
with negative magnetization, is stabilized in its place.

The coreless vortex first splits into the pair of singular vortices
with polar cores, such that the
doubly quantized vortex in $\zeta_-$ splits into two singly quantized
vortices. As the energy relaxes,
$\zeta_+$ is depopulated leaving a spinor with winding numbers $1$ and
$2$ in $\zeta_0$ and $\zeta_-$
respectively. The vortices in $\zeta_-$ then exit the cloud, leaving a
spinor of the same form as that
for strong, positive magnetization,
with $\zeta_\pm$ interchanged. The structure is as described above
with the spin rotated by $\pi$ about
an arbitrary axis in the $x$-$y$ plane.

\subsubsection{Polar Regime}
\label{sec:coreless-polar-regime}

Next we study the energetic stability of a FM
coreless vortex in the polar interaction regime.
In spin-1 BECs of $^{23}$Na, which exhibit polar interactions, FM
coreless vortices have been prepared experimentally
via magnetic field rotation~\cite{leanhardt_prl_2003,choi_njp_2012}.
Simple energetic arguments would predict that the coreless vortex is
then always unstable.  However, conservation of magnetization requires
that the condensate always exhibit non-polar regions, which we will
show allows the coreless vortex to remain stable.
As an initial state we take the experimentally phase-imprinted state
[Eq.~\eqref{eq:cl}] with $F=1$ everywhere, for different $M$.
The energetic stability and structure of the vortex is then determined
by numerically minimizing the free energy
in a rotating trap (at the frequency $\Omega$).

In a spin-1 BEC a vortex singularity can be accommodated by exciting the
wave function out of its ground-state manifold, whenever it is energetically
more favorable to adjust the spin value than force the density to vanish at the
singular core~\cite{ruostekoski_prl_2003,lovegrove_pra_2012}. This
happens
when the characteristic length scales
$\xin$ and $\xiF$ [Eq.~\eqref{eq:healinglengths}] satisfy $\xiF\agt\xin$.
In addition, conservation of magnetization now
introduces $\etaM$ [Eq.~\eqref{eq:magnetization-healing-length}] as a
third length scale that describes, in an otherwise polar condensate,
the size of a FM core needed to yield a specific $M$.

\begin{figure}[tb]
  \begin{center}
    \includegraphics[width=\columnwidth]{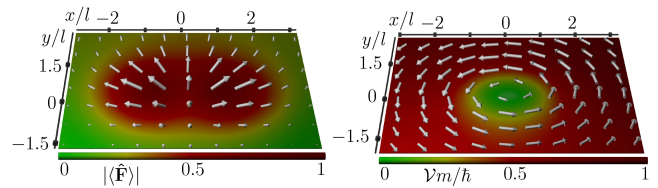}
  \end{center}
  \caption{(Left) Spin profile $\inleva{\F}$ (arrows) and $\inlabsF$ (color
    gradient and arrow lengths)
    of the coreless vortex in the polar regime, interpolating between
    FM and polar phases and displaying
    the characteristic fountain texture inside the core of a singular
    polar vortex.
    (Right) The corresponding superfluid velocity $\mathbf{v}$ and its
    magnitude (arrows)
    and circulation density $\mathcal{V}=\rho\mathbf{v}\cdot\phihat$ (color
    gradient),
   continuously interpolating from nonsingular to singular circulation.
  }
  \label{fig:polar-cl}
\end{figure}

As the energy of the imprinted coreless vortex relaxes, the
spin-dependent interaction drives the condensate towards the polar
phase with $F\rightarrow0$ away from the vortex line. The outer
region then approaches the singly quantized vortex~\eqref{eq:p012},
exhibiting a radial disgyration of $\nematic$.
In the limit of weak magnetization, the vortex core splits into two
half-quantum vortices, similarly to the splitting of a
singly quantized vortex predicted when magnetization is not
conserved~\cite{lovegrove_pra_2012} and observed in
experiment~\cite{seo_prl_2015}.
At $M=0$, the fountain texture is lost entirely. When $M$ increases,
the stable coreless-vortex spin texture gradually becomes more
pronounced, preventing the core splitting.
The vortex, shown in Fig.~\ref{fig:polar-cl},
still exhibits axial
asymmetry in the magnetized core region, with two close-lying
spin maxima $F=1$, and $\inleva{\F} \parallel \zhat$ at the
center. Ignoring the slight core asymmetry, the vortex can be
qualitatively described by the analytic model~\eqref{eq:cl-p012}: The spin
winds to $\inleva{\F} \parallel \rhohat$ as $\rho$
increases. Simultaneously, $F$ decreases sharply and the configuration
approaches a singly quantized polar vortex.
The size of the core (the magnetization density half width at half
maximum) is $\sim\etaM$.
Comparison of length scales then suggests that the coreless texture
becomes pronounced
when $\etaM\agt \xiF$ ($>\xin$), which is in qualitative agreement
with our numerical results.

Owing to the trap, $F$ can reach a local minimum---which may not vanish
in all directions---and start increasing at the edge of the
cloud. The vortex profile then depends on $M$ and additionally, e.g., on the
quadratic Zeeman shift, which favors the polar
phase~\cite{ruostekoski_pra_2007}. We may therefore
envisage a scheme to engineer the core symmetry and even more complex
composite defects by Laguerre-Gaussian lasers that generate a Zeeman
shift with a cylindrical shell symmetry, which we explore further in
section~\ref{sec:composite-core}.

It follows from Eq.~\eqref{eq:nu-winding}
that circulation alone is quantized only in the polar phase ($F=0$).
Figure~\ref{fig:polar-cl} demonstrates how the superfluid circulation
in the relaxed vortex state smoothly connects the small- and
large-distance topology, as qualitatively described by
Eqs.~\eqref{eq:cl-v}--\eqref{eq:circ-dens} for the 
analytic model~\eqref{eq:cl-p012}.

We note, finally, that also in the polar interaction regime the
three-component coreless vortex gives way to the effective
two-component state in Fig.~\ref{fig:two-comp} when magnetization is
sufficiently strong.  
The transition to the two-component system occurs when the
$\beta(\rho)$ profile no longer allows the three-component
vortex~\eqref{eq:cl-p012} to satisfy the magnetization constraint,
depopulating $\zeta_-$. The threshold magnetization value decreases
with rotation (Fig.~\ref{fig:stability}). At the lowest frequency
where the coreless vortex is stable, the two-component regime appears
at $M \simeq 0.25$. When the rotation is nearly rapid enough to  
nucleate additional vortices, the threshold magnetization has decreased
to $M \simeq 0.15$.
For large $c_0$, the stable two-component
solution represents a smooth transition from the $F=1$ coreless vortex
at the trap center to the $F=0$ singular vortex at the edge of the
cloud. 

\subsection{Nematic Coreless Vortex}
\label{sec:ncv}

When the magnetic-field rotation technique used to phase-imprint the FM
coreless vortex~\cite{leanhardt_prl_2003,choi_njp_2012} is
applied to a BEC prepared initially in the state $\zeta =
(0,1,0)^T$, representing the polar phase with $\nematic=\zhat$
and longitudinal magnetization $M=0$, the result is
a nematic coreless vortex~\cite{choi_njp_2012,choi_prl_2012},
Eq.~\eqref{eq:ncv}. This is
characterized by a fountainlike texture
$\nematic = \sin\beta^\prime\rhohat + \cos\beta^\prime\zhat$ of the
nematic axis. Since
the condensate is unmagnetized prior to the magnetic-field rotation, the
longitudinal magnetization remains zero in the imprinted texture.
This $\nematic$ texture can continuously unwind to the
uniform state and has vanishing mass circulation. It can therefore not
be stabilized by rotation as the coreless vortex can.

We ask instead whether the nematic coreless vortex can be stable
inside the core of
a composite topological defect when a conserved, nonzero magnetization
necessitates the
formation of non-polar regions.
A nematic coreless vortex with a nonzero magnetization could be created by
phase-imprinting via population
transfer~\cite{matthews_prl_1999,andersen_prl_2006,leslie_prl_2009}
that individually prepares
the appropriate phase windings of $\wno{-2\pi}{0}{2\pi}$ in the spinor
components.
In a magnetized BEC, $F$ will acquire a spatial structure
interpolating between $F=0$ at the center of the cloud to $F>0$
at the edge as energy relaxes. This behavior is described in the
analytic model in Eq.~\eqref{eq:nematic-vortex}.
From $\nematic\perp\inleva{\F}$,
it follows that in order to have the fountain texture in $\nematic$ we
must have $\beta=\pi/2$ at $\rho=0$ and increasing monotonically.
The corresponding mass circulation,
$\oint d\mathbf{r} \cdot \mathbf{v} =-\frac{hF}{m}\cos\beta$,
interpolates from the non-circulating polar core to a nonzero
circulation, allowing, in principle, stabilization by
rotation.

\begin{figure}[tb]
  \begin{center}
    \includegraphics[width=\columnwidth]{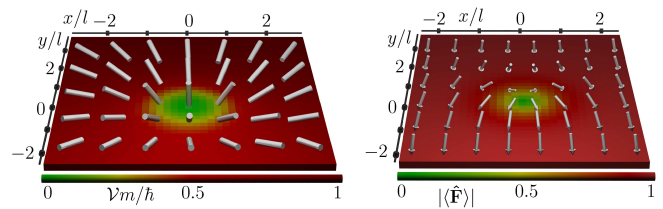}
  \end{center}
  \caption{Stable nematic coreless
    vortex in a BEC with polar interactions. (Left) The unoriented
    $\nematic$-vector
    (cylinders) exhibits the coreless fountain-like texture. The
    circulation density $\mathcal{V}=\rho\mathbf{v}\cdot\phihat$ (color
    gradient) shows the
    composite-vortex structure, interpolating between the
    non-circulating polar phase to the outer singly quantized FM
    vortex.
    (Right) Corresponding spin texture $\inleva{\F}$ (arrows) and spin
    magnitude $\inlabsF$ (color gradient and arrow lengths), showing the
    core region.  Conservation of
    magnetization forces the BEC into the FM phase away from the
    vortex line.
  }
  \label{fig:nematic-cl}
\end{figure}

For sufficiently strong
magnetization the condensate will reach $F=1$ in an outer FM region. 
This outer,
FM region represents the large-distance topology of a singular FM
vortex, while the topology at small distances is represented by the nematic
coreless vortex.  We find that the numerical relaxation process of the
phase-imprinted $\wno{-2\pi}{0}{2\pi}$ vortex configuration at given
$M$ can be qualitatively described by the spinor~\eqref{eq:nematic-vortex}.
However,
our numerics demonstrate that the
vortex is energetically stable only once magnetization is strong
enough to deplete $\zeta_+$, enforcing an effective two-component
regime. We find this to occur at $M\lesssim-0.2$. The stable vortex
state shown in Fig.~\ref{fig:nematic-cl}, then
exhibits a Mermin-Ho-like texture in $\nematic$ [corresponding to a charge
$W=1/2$ as defined by Eq.~\eqref{eq:cl-W} with $\nhat_F=\nematic$],
and a corresponding bending
of the spin vector from the $\rhohat$ direction at the center to the
$-\zhat$ direction in the FM region. The core size is again determined by
the magnetization constraint.  
(We note in Sec.~\ref{sec:fm-weak} that relaxing the energy of a
singly quantized FM vortex also results in an effective two-component
nematic coreless vortex when magnetization is sufficiently strong,
demonstrating that the nematic coreless vortex can be
stable also in the FM interaction regime.) 
The instability of the nematic coreless vortex
at weaker magnetization results from the existence of
lower-energy singular vortices with FM cores.

\section{Singular vortices}
\label{sec:weak-mag}

In Ref.~\cite{lovegrove_pra_2012}
we calculated the core structures of energetically stable singly quantized
vortices in the spin-1 BEC. In that analysis,
the magnetization
was allowed to vary during relaxation. We now calculate how these
structures change if a weak longitudinal magnetization is preserved
throughout the relaxation procedure.  We find that the structure of
the vortex cores is not notably affected unless the value of
magnetization becomes very large. When the magnetization exceeds a
critical value, the condensate relaxes to an effective two-component
regime.

\subsection{Singular FM vortex}
\label{sec:fm-weak}

For this analysis, it is instructive first to review the central
features of the state that results from energy minimization
if conservation of magnetization is not imposed. A trial wave
function representing a singly quantized FM vortex is given by
Eq.~(\ref{eq:fmsingular}). The vortex is made up
of overlapping vortex lines in the three spinor components,
corresponding to a depletion of the atom density in the vortex core.
The spin texture is uniform. As the energy
is relaxed, these vortex lines move apart, such that the atom density
is nonzero everywhere. The FM order parameter, however, remains
singular on a well-defined vortex line, on which the atoms adopt the
polar phase.

The size of a density-depleted vortex core would be determined by the
density-healing length $\xin$ [Eq.~(\ref{eq:healinglengths})]. By allowing the
core to fill by perturbing $\absF$, the core can expand to the size of
the spin healing length $\xiF$ [Eq.~(\ref{eq:healinglengths})], thereby lowering
its energy.
The deformation of the vortex core corresponds to a local rotation of the
spin vector in an extended core region. The result is a spin winding
around a core with nonvanishing
density and $\absF < 1$. Away from the vortex core the initial uniform spin
texture is preserved.

The stability of the singular FM vortex seems counter-intuitive, since
there also exists a stable 
coreless vortex with lower energy for the same parameters. However, in
order for the singular vortex to decay into a coreless vortex with
lower energy, the singular vortex would first have to exit the cloud
(which requires a relatively slow rotation frequency), followed by nucleation
of the coreless vortex (requiring sufficiently rapid rotation to overcome
the energy barrier associated with vortex nucleation). The FM singly
quantized vortex is therefore stable as a local
energy minimum for a range of rotation frequencies.

In Ref.~\cite{lovegrove_pra_2012} we demonstrated how analysis and
classification of the vortex core structure is facilitated by a basis
transformation to a \emph{natural spinor basis} for the vortex state.
In the absence of external magnetic fields, we are free to choose the
spinor basis.

By transforming to the basis of spin
projection onto the axis defined by the uniform
$\inleva{\mathbf{\hat{F}}}$ far
from the vortex core, the spinor
representing the relaxed core can be written in the
form of an interpolation between an outer singular FM
vortex and inner noncirculating polar phase, similar to
Eq.~(\ref{eq:nematic-vortex}). In this natural
basis, the spinor reads
\begin{equation}
  \label{eq:fm-natural}
  \zetaN=\frac{1}{2}\threevec{\sqrt{2} e^{i \varphi}\left(\cos^2\frac{\betaN}{2}D_+- \sin^2\frac{\betaN}{2}D_-\right)}
  {\sin\betaN\left( D_+  + D_-\right)}
  {\sqrt{2}e^{-i \varphi}\left( \sin^2\frac{\betaN}{2}D_+- \cos^2\frac{\betaN}{2}D_-\right)},
\end{equation}
where the angle $\betaN(\rho)$ describes the tilt
of the spin away from the new quantization axis.
The vortex lines in $\zetaN_\pm$ overlap and the core is filled by
$\zetaN_0$.
The interpolation between the FM vortex and the polar
core is described by $D_\pm(\rho) = \sqrt{1-F(\rho)}$ as defined in
the construction of Eq.~\eqref{eq:general}.
The energetically stable vortex has the large-distance topology of a
singular FM vortex, exhibiting a radial spin
disgyration close to the vortex core. In the core of size
$\sim\xiF$, the topology represents the vortex-free polar phase.

In this natural basis we may define a magnetization
\begin{equation}
  \label{eq:Mstar}
  \MN=(\NN_{+} - \NN_{-})/N,
\end{equation}
where $\NN_\pm$ are the
populations of the $\zetaN_\pm$ spinor components. In the
initial state, $\MN=1$.
Owing to the rotation of the spin vector around
the vortex core in the relaxed state, the  contribution of the core
region to the net 
magnetization is canceled out. Therefore $\MN$ decreases during the
relaxation. The spin then bends to point in the direction of the
natural spinor basis over a non-negligible distance from the vortex
core, leading to a further reduction in $\MN$.
In the relaxed state, $\MN \simeq 0.5$ at the lowest
rotation frequency where the
vortex becomes stable, increasing to $\MN \simeq 0.7$ at
the upper limit of stability.

\begin{figure}[tb]
  \centering
  \includegraphics[width=\columnwidth]{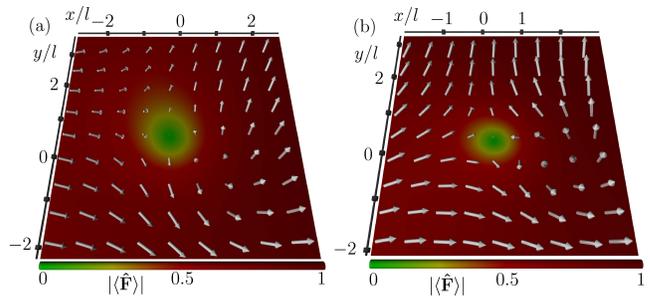}
  \caption{Numerically calculated spin vector (arrows)
    and magnitudes
    (color gradient) for the singular vortex of the FM phase in the
    $x$-$y$ plane with
    a conserved magnetization of (a) $M=0.2$, (b) $M=0.5$. Arrow lengths
    scale with
    the spin magnitude. The core structure remains
    as described in Ref.~\cite{lovegrove_pra_2012}, with the conserved
    magnetization
    leading to a local rotation in spin space as the vortex core deforms.
  }
  \label{fig:fm-singular-weak}
\end{figure}

We are now in a position to understand how conserving an initial longitudinal
magnetization $M$, defined by
Eq.~(\ref{eq:M}) in the
basis of spin projection onto the $z$ axis, changes the relaxed state.
The trial wave function
for the singular FM vortex is again given by
Eq.~(\ref{eq:fmsingular}), where the constant angle $\beta_0$ is now chosen
to yield the desired $M$. Energy relaxation under the condition that
$M$ is conserved now results in the spin structure shown in
Fig.~\ref{fig:fm-singular-weak}. Again relaxation results in
a local rotation of the spin vector to allow the vortex core to avoid the
density depletion and instead fill with atoms with $\absF<1$,
expanding the size of the core to $\xiF$ as illustrated in
Fig.~\ref{fig:fm-singular-weak}. However, when the magnetization is
conserved, in addition to $\xin$ and $\xiF$, also the magnetization
length scale $\etaMs$ given by Eq.~\eqref{eq:maglength-polcore}
may affect the core structure. The length $\etaMs$ defines the upper
limit on the size of the non-FM core for any given $M$. As long as $M$
is sufficiently weak that $\etaMs \gtrsim \xiF$,
the core size after energy relaxation is determined by $\xiF$, and
the filling of the vortex core may be understood from minimization of
the gradient energy.

While the general understanding of the core-deformation mechanism is
not qualitatively changed as long as $M$ is sufficiently weak, the
resulting spin texture must adapt to the conserved magnetization. The
general mechanism whereby the structure of the vortex core emerges as
a result of the energetics 
associated with the two characteristic length scales $\xin$ and
$\xiF$ thus remains unaffected when conservation of magnetization is
imposed. This description holds until $\xiF>\etaMs$, at which
point the magnetization length constrains the core size.

Compared with the result found when not accounting for
conservation of magnetization, the spins are everywhere tilted towards
the $z$ axis. This compensates for the amount of magnetization that
would otherwise be lost in the formation of the core region. Otherwise,
the core structure remains qualitatively unchanged. Increasing $M$
leads to a further local rotation of the spin vector
everywhere towards the $z$ axis, as shown in Fig.~\ref{fig:fm-singular-weak}.
We may then conclude that the effect of conserving a fixed longitudinal
magnetization is to fix the natural basis of the
vortex as long as $\abs{M} \leq \MN$. Greater longitudinal
magnetization cannot be achieved by tilting of the spin structure
shown in Fig.~\ref{fig:fm-singular-weak}.
Our analysis thus immediately predicts a maximum
magnetization $\abs{M}=\MN$ above which the vortex state has to
change, as stronger magnetization cannot be provided. At this
magnetization strength, also $\etaMs \sim \xiF$ in our
simulations, implying that the size of the vortex core becomes
determined by the magnetization length scale rather than the spin
healing length. 
At this point, magnetization causes depletion of the minority
component, leading to an effective two-component vortex whose core
region exhibits a nematic-coreless-vortex texture.

\subsection{Polar half-quantum vortex}
\label{sec:polar-weak}

We now consider a condensate with a single
half-quantum vortex. It is instructive first to ignore
conservation of magnetization and analyze the resulting relaxed state.
A trial wave function carrying a
half-quantum vortex may be constructed from Eq.~(\ref{eq:hq}) by applying
a spin rotation such that all spinor components have a
nonzero population.

The trial wave function corresponds to a vortex where the atomic
density vanishes on the singularity. The size of the core is then
determined by the density healing length $\xin$
[Eq.~(\ref{eq:healinglengths})]. As energy relaxes, the vortex core is
filled with
atoms with $\absF>0$,
reaching the FM phase on the singularity of the polar order
parameter. The vortex core can then expand to the size of the spin
healing length $\xiF$ [Eq.~(\ref{eq:healinglengths})]. In addition, a small
region of nonzero $\absF$, forms near
the edge of the condensate, in which the spins anti-align with the
spin inside the vortex core. This effect appears counter-intuitive,
as exciting the wave function out of the polar phase costs interaction
energy. This cost is, however, relatively small in the low-density
region of the cloud, and is offset by lowering the gradient
energy arising from the filled vortex core.

Similarly to the analysis of
the singular FM vortex, we may find a natural basis by transforming
the wave function to the basis of spin quantization along the axis
defined by the spin vector on the vortex line. The spinor then reads
[cf.\ Eqs.~\eqref{eq:hq} and \eqref{eq:fm-hq2})],
\begin{equation}
\label{eq:hq_mag}
\zetaN^{\rm hq}= \frac{1}{\sqrt{2}}
    \threevec{-\sqrt{1+ \MMN(\rho)}}
             {0}
             {e^{i \varphi} \sqrt{1- \MMN(\rho)}}\,,
\end{equation}
where the local magnetization $\MMN$ describes the filling of the
vortex core and the magnetization of the cloud edge. The large-distance
topology of this vortex is that of the half-quantum vortex with nonzero
magnetization (note that the FM phase is not observed far from the core)
and the small-distance topology is that of the vortex-free FM phase.

With these observations in mind, we can now analyze the consequences
of preserving a nonzero longitudinal magnetization of the
vortex-carrying condensate.
In order to give the trial wave function representing the half-quantum
vortex a nonzero magnetization, we renormalize the occupations
of the spinor components.

\begin{figure}[tb]
  \centering
\includegraphics[width=\columnwidth]{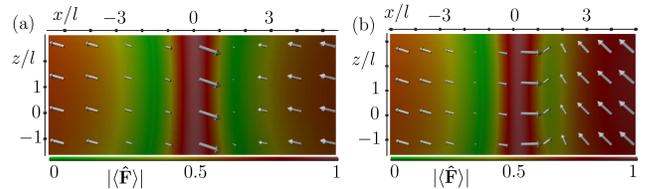}
  \caption{Spin vector (arrows) and magnitude (color gradient)
  profiles in the $x$-$z$ plane for a half-quantum vortex with an
  initial magnetization $M=0.2$ 
  which is (a) not conserved and
  (b) conserved. Arrow lengths scale with spin magnitude. The
  magnetization arises from the outer regions, not the FM core.
  }
  \label{fig:pol-half-quantum}
\end{figure}

The relaxed half-quantum vortex state with fixed magnetization is
shown in Fig.~\ref{fig:pol-half-quantum}. On the vortex line,
$\absF=1$, and the core lowers its energy by expanding to the size
allowed by the spin healing length. We also note that a significant
region with nonzero $\absF$ arises towards the edge of the cloud.
From the spin texture we note that the longitudinal magnetization
$M>0$ arises not from the vortex core, but from the magnetized edge
regions. The spins in the edge regions
remain nonzero to reduce spin gradients. However, the spins in
the edge regions no longer anti-align with that inside the vortex core.
This means that the treatment in terms of a natural spinor basis is no longer
valid in a magnetized half-quantum vortex. The effect of fixing a weak
longitudinal magnetization is to increase the magnitude of the spin in the
outer region and to orient the spin in this region towards the direction
of the applied magnetic field.

We find that magnetization of the
edge region can only provide a total magnetization of $\abs{M} \sim
0.3$, beyond which the vortex is no longer energetically
stable. This is approximately the magnetization at which
$\etaMs<2\xiF$, at which point the half-quantum vortex
carries a large gradient energy.
However, the half-quantum vortex can
be stabilized at greater magnetization, by a negative quadratic
Zeeman splitting which is sufficiently strong to overcome the
gradient energy. With $g_2 B_z^2 = -0.2\hbar\omega$ in
Eq.~(\ref{eq:hamiltonian-density}) the half-quantum vortex remains
stable up to $\abs M \sim 0.8$.

\subsection{Polar singly quantized vortex}

The energetic stability and structure of a singly quantized vortex in
a polar spin-1 BEC were analyzed in detail in
Ref.~\cite{lovegrove_pra_2012}. In that analysis, the initial state
was entirely in the polar phase, and the magnetization was allowed to
vary during energy relaxation. Energy minimization then resulted in a
splitting of the singly quantized vortex into a pair of half-quantum
vortices with FM cores, whose spins anti-aligned.  In experiment, the
core spin polarization was used to image the half-quantum vortices and
thereby observe the splitting process~\cite{seo_prl_2015}. This splitting
preserves the overall topology of the initial state, but forms an
extended core region where the phases mix. The filling of the vortex
cores with atoms with $\absF\ge0$, and accommodating the singularities
by requiring $\absF=1$ on the singular lines, lowers the total
energy by reducing the gradient energies.

\begin{figure}[tb]
  \centering
  \includegraphics[width=\columnwidth]{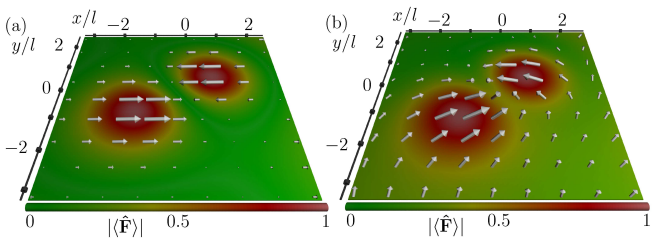}
  \caption{Numerically calculated spin magnitude (color
    gradient) and
    vector (arrows) profiles of the singular polar vortex in the $x$-$y$
    plane when an initial magnetization $M=0.2$ is (a) not conserved and
    (b) conserved. Arrow lengths scale with spin magnitude. The singular
    vortex deforms to a pair of half-quantum vortices with FM cores in
    both cases.
  }
  \label{fig:pol-singular}
\end{figure}

Similarly to the cloud with a single half-quantum
vortex, the gradient energies associated with the vortex cores are
again lowered by the formation of two magnetized edge regions,
illustrated in Fig.~\ref{fig:pol-singular}(a). The
spins in the two edge regions anti-align with those in the
nearest vortex core, immediately implying that the two edge regions
exhibit spins pointing in opposite directions.

The formation of the magnetized edge regions is reminiscent of a
composite topological defect
such as those described by Eqs.~(\ref{eq:101-general}) and
(\ref{eq:phasevortex-general}). However,
the breaking of axisymmetry renders the interpretation of this
structure in terms of a hierarchical core structure
prohibitively difficult.

This picture remains unchanged as we account for a conserved
magnetization of zero. This is to be expected as the structure described above
has zero net magnetization. However, a conserved, weak, nonzero
magnetization does modify the structure somewhat. A magnetized singly quantized
vortex can be constructed from
Eq.~(\ref{eq:psingular}) by adjusting the populations of the spinor
components to give the desired magnetization. We now explicitly conserve this
magnetization throughout energy relaxation. The resulting relaxed
states for conserved and nonconserved weak, nonzero magnetization are
shown in Fig.~\ref{fig:pol-singular},
and we note the splitting of
the initial vortex into two singular lines in both cases, which may be
identified as half-quantum vortices.

In contrast to the isolated half-quantum vortex, the nonzero magnetization $M$
of the condensate is carried by the vortex cores rather than the edge
regions. When the condensate contained
only a single half-quantum vortex, the spin structure could adjust to
a varying magnetization $M$ by a simple rotation of the spins everywhere in
the cloud. Because the spins in the two
edge regions anti-align when magnetization is not conserved, an equal
rotation of spins everywhere can no
longer increase the longitudinal magnetization. Therefore, the effect
of preserving a nonzero $M$ throughout energy
relaxation is to cause the spins in the two vortex cores to orient
differently so that 
they no longer anti-align. This provides the required net
magnetization. The gradient energy in the extended core region
is reduced by this process and so the magnitude of the spin in
the edge regions is not required to be as strong as in the case
of zero magnetization.

Increasing the magnetization causes the spins in the magnetized cores
to orient toward the
direction of the applied magnetic field. The cores also draw closer
together. Further increasing
the magnetization leads to the non-polar regions overlapping and
forming a single core with a
continuous spin texture, exhibiting the fountain structure associated
with the coreless vortex.  Coexistence of different core symmetries of
the singly quantized polar vortex is explored in
Ref.~\cite{borgh_prl_2016}.  There it is shown how the core isotropy is
sensitive to tunable Zeeman shifts, which can continuously restore the
axial symmetry.  Here we explain how the vortex core structure arises
from the magnetization constraint that the FM core of the vortex line
can provide.

The core regions can only provide a weak magnetization as their size is
constrained by the spin healing length. The magnetization length
scale $\etaM$ given by Eq.~\eqref{eq:magnetization-healing-length}
described the 
smallest core size required to yield a given magnetization $M$.
Hence, when $M$ is sufficiently strong that $\etaM \gtrsim \xiF$,
the magnetization cannot be upheld by the vortex cores alone, leading to
energetic instability of the state in
Fig.~\ref{fig:pol-singular}. This leads to depletion of the minority
component, resulting in a two-component coreless vortex similar to
Fig.~\ref{fig:two-comp}.
We find that this happens at $\inlabs M \sim 0.3$.
While a negative
quadratic Zeeman splitting could restabilize a single half-quantum vortex at
higher magnetization, the same is not true for the split singly
quantized vortex.

In conclusion, we find that also in the polar regime, the results of
Ref.~\cite{lovegrove_pra_2012} remain qualitatively unchanged when
accounting for conservation of a sufficiently weak initial
magnetization, exhibiting the splitting of the vortex core. However,
the required magnetization in the relaxed state is produced by forcing
the resulting FM vortex cores to no longer exhibit anti-aligning
spins. As magnetization increases, the vortex cores gradually merge,
eventually exhibiting an effective two-component coreless vortex as $\inlabs{M}
\gtrsim 0.3$. Hence, in the polar interaction regime, initial 
coreless and singly quantized vortices relax to the same
defect structure at both weak and strong magnetization.

\section{Examples of complex vortex-core hierarchies}
\label{sec:composite-core}

In this section we consider the more complex composite vortices
implied by the solutions in Table~\ref{tab:analytics} that connect
singular vortex states in both their FM and polar limits.  In
these cases a more complex hierarchy could form, exhibiting several
concentric core regions where the phases alternate as the inner
singular vortex itself develops a filled core. These highly complex
core structures 
are not in general energetically stable.  However, we demonstrate here
how FM vortices with polar half-quantum or singly quantized vortex
small-distance topology may be stabilized.

\subsection{Half-quantum vortex core of FM vortex}

In Table~\ref{tab:analytics}, we presented analytic vortex solutions
representing composite topological defects, including the nontrivial
case of a singular polar vortex forming the core of an outer singular
FM vortex. [See also Eq.~\eqref{eq:fm-hq2}].
As energy
relaxes, a three-step vortex hierarchy may form: The large-distance
topology represents the singular FM vortex with
spin $\expF=\sin\beta\xhat+\cos\beta\zhat$. Inside its core,
$F(\rho)$ decreases to $0$, displaying a half-quantum vortex with
$\nematic = \cos\beta\cos(\varphi/2)\xhat + \sin(\varphi/2)\yhat -
\sin\beta\cos(\varphi/2)\zhat$. 
To avoid depletion of the atom density, $F(\rho)$ may then increase
back to $F(\rho\to0)=1$ inside the core
of the half-quantum vortex, corresponding to the vortex-free FM
phase.
A similar composite vortex state was considered in
a recent topological classifications of vortex cores in spin-1
BECs~\cite{kobayashi_pra_2012}.

We now ask whether the composite-vortex structure can form as the
energy of the singular FM vortex relaxes, and whether it can be
energetically stable.
We consider again the trial wave function for a singly quantized FM
vortex, constructed from Eq.~(\ref{eq:fmsingular}) with magnetization
$0\leq M\leq0.8$, in a condensate with FM interactions. We find
that in order for the composite-vortex structure to replace the
vortex-free polar core in the stable state, a sufficiently strong
negative quadratic Zeeman splitting is required. This can be induced
by combining a static magnetic
field with a microwave dressing field, generating and AC Stark shift
that corresponds to a highly tunable quadratic level
shift~\cite{gerbier_pra_2006}. The level shift may also be induced by
lasers~\cite{santos_pra_2007}.
Here we take $g_2
B_z^2 = -0.2\hbar\omega$.
The negative quadratic Zeeman effect favors occupation of the
$m=\pm 1$ Zeeman levels. This causes the spin vector to align
(anti-align) with the $z$ axis away from the vortex, and to
anti-align (align) with it in the FM core. The two possible spin
alignments are energetically degenerate (though conservation of
magnetization may only allow one).

The quadratic Zeeman splitting required to energetically stabilize the
composite-vortex structure is strong enough that  $\beta$ is forced to
adopt values of either $0$ or $\pi$, and the $\zeta_0$ spinor component
is empty. In the resulting effective two-component limit, the spinor
can be parametrized as
\begin{equation}
\label{eq:sdv}
\zeta^{\rm c}= \frac{1}{\sqrt{2}}
    \threevec{-e^{i \varphi}\sqrt{1+F_z(\rho)}}
             {0}
             {\sqrt{1-F_z(\rho)}}\,,
\end{equation}
where $\expF=F_z\zhat$ with $F_z=-1$ in the
inner core, and $F_z=1$ away from the vortex.
It is now readily apparent that the vortex line in $\zeta^{\rm c}_+$
represents the overall topology of the singly quantized FM
vortex at sufficiently large $\rho$.
Similarly in the inner core, $\zeta^{\rm c}_-$
represents a vortex-free FM wavefunction. Where $F_z\to0$ in the
intermediate region, $\zeta^{\rm c}$ takes the form of a
half-quantum vortex similar to Eq.~(\ref{eq:hq}).

Due to the FM interaction, the thickness
of the polar region is restricted by $\xiF$. Hence the
polar vortex takes on the character of a domain wall separating an outer
spin domain from an inner domain with opposing spin. The size of the inner
FM core is not constrained by the spin healing length, as it does not
violate the FM spin condition.

\begin{figure}[tb]
  \centering
  \includegraphics[width=\columnwidth]{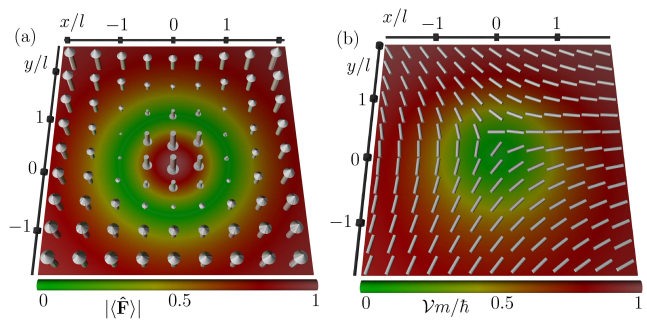}
  \caption{(a) Numerically
    calculated spin texture (arrows) and magnitude
    (color gradient) in the $x$-$y$ plane for the composite vortex in
    the FM regime with
    $M=0.6$. Arrow lengths scale with spin magnitude. (b) Nematic axis
    profile (cylinders)
    and circulation (color gradient) for the same, showing the
    non-circulating inner FM core
    and winding of $\nematic$ by $\pi$.
  }
  \label{fig:composite}
\end{figure}

However, in the atomic spinor BEC, magnetization is conserved. The
effect of this is to determine the size of the inner
FM core, so that the antialigned spins in the FM regions yield the required
magnetization. This corresponds to a length scale $\etaMd$
associated with the conserved magnetization. In a simplified model
that ignores the thickness of the domain wall,
an estimate for $\etaMd \simeq R_\mathrm{TF} \sqrt{1-[(1+M)/2]^{2/5}}$
may be derived analogously to the 
magnetization length scales defined in section~\ref{sec:mft}.
From this we can understand the
upper limit on magnetization for which the composite vortex is
stable: The magnetization must not cause the gradient energies
associated with a small core to overcome the Zeeman energy. This
happens when $\etaMd \lesssim \xiZ =
l(\hbar\omega/2|g_2|B_z^2)^{1/2}$, the healing length associated with
the quadratic Zeeman energy, in agreement with our numerical results.
The vortex structure at $M=0.6$ is shown in Fig.~\ref{fig:composite},
demonstrating the formation of the composite core in which the innermost
region exhibits the noncirculating FM phase.

It is interesting to note that the conservation of a nonzero magnetization
is required to stabilize this vortex in the FM regime. If the magnetization
is not conserved, or is conserved at zero, the gradient energy associated with
the domain wall renders the vortex unstable.

A sufficiently strong magnetization $\abs{M} \gtrsim 0.3$ together
with the quadratic Zeeman energy also allows the composite-vortex
structure to be energetically stable in a BEC with polar
interactions. The strong magnetization is upheld by forcing the
condensate away from the vortex line towards the FM phase, forming a
singular vortex. Again the required Zeeman energy causes depopulation
of $\zeta_0$. In contrast to the FM regime, the polar interactions
now imply that that size of the FM core is determined by the spin
healing length, while the thickness of the polar region is determined
by the magnetization. The quadratic Zeeman
splitting is able to stabilize the composite vortex provided
$\etaMd>\xiZ$, corresponding to magnetization of $M\sim0.9$,
which is supported by our numerics. When magnetization is not conserved,
both the half-quantum and singly quantized vortices of the polar phase
described in Sec.~\ref{sec:polar-weak} are stable under the influence of
quadratic Zeeman splitting. For magnetization $M<0.3$, the half-quantum
vortex remains stable. The stronger magnetization serves to increase $\absF$
near the edge of the trap, forming the FM phase.

\subsection{Singly quantized vortex core of FM vortex}

Inspired by the results described in Fig.~\ref{fig:composite} and the
observations made in section~\ref{sec:coreless-polar-regime}, we can
envisage using a designed quadratic Zeeman shift to stabilize complex
composite defect structures by locally forcing the condensate into the
polar phase.  A highly tunable quadratic level shift
can in principle be imposed using lasers~\cite{santos_pra_2007}.
Consider, e.g., the spatial profile of a Laguerre-Gaussian beam,
\begin{equation}
  \label{eq:lg-zeeman}
  q = q_0\left(\frac{2\rho^2}{w^4}\right)^m
      e^{-2\rho^2/w^2}\left[L^m_n\left(\frac{2\rho^2}{w^2}\right)\right]^2,
\end{equation}
where $m$ and $n$ give the Laguerre-Gaussian modes and $w$ is the
beam-waist parameter. 
If a quadratic Zeeman shift is induced using Eq.~\eqref{eq:lg-zeeman},
concentric regions can be created where the polar (FM) phase is
favored by the presence (absence) of the 
level shift. Using high Laguerre-Gaussian modes, these regions may be made quite
sharp.

\begin{figure}[tb]
  \centering
  \includegraphics[width=\columnwidth]{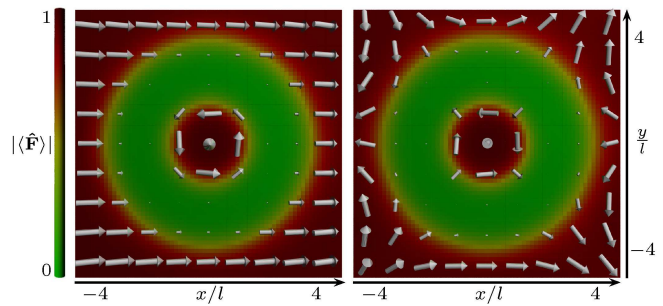}
  \caption{Spin magnitude (color scale) and spin vector
    (arrows) showing composite vortex states arising as a singly
    quantized vortex in a FM condensate relaxes in the presence of an
    engineered quadratic Zeeman shift with Laguerre-Gaussian profile
    that forces the 
    condensate into the polar phase in an annulus surrounding the
    vortex core. (Left) Trap rotation $\Omega=0.125\omega$. A coreless
    vortex appears in the inner FM core, surrounded by a singly
    quantized polar vortex, together forming the extended core of a
    singular FM vortex. (Right) $\Omega=0.135\omega$. Rotation causes a
    multiply quantized vortex to form in the outer FM region.
  }
  \label{fig:lg-zeeman}
\end{figure}

As an illustration of a composite vortex structure stabilized this way,
we
consider a FM condensate containing a phase-imprinted singular vortex
at $M=0$. This is allowed to relax, conserving $M$ throughout, in the
presence of a positive quadratic Zeeman shift with $(m,n)=(6,0)$
Laguerre-Gaussian profile. (In order to more easily produce a sharply
defined vortex 
structure we assume $Nc_0=10^4\hbar\omega l^3$ together with an
artificially strong FM interaction $Nc_2=463\hbar\omega l^3$.)
The resulting vortex state is shown in Fig.~\ref{fig:lg-zeeman} for
two different trap-rotation frequencies.  The positive quadratic
Zeeman shift favors the $\zeta_0$ component and locally forces the
condensate into the polar phase in a ring-shaped region.  
The relaxed vortex state exhibits a concentric core structure, where
an inner FM coreless vortex is surrounded by a singly quantized polar
vortex, which in turn is surrounded by the singly quantized FM vortex
far away from the vortex line.
If rotation is increased, the part of the cloud corresponding to the
large-distance topology may exhibit a multiply quantized vortex. 
Such a solution is shown to the right in Fig.~\ref{fig:lg-zeeman}.

\section{Concluding remarks}
\label{sec:conclusions}

On time scales where $s$-wave scattering dominates the
interactions in a spin-1 BEC, the longitudinal magnetization
$M = (N_+-N_-)/N$ is approximately
conserved~\cite{jacob_pra_2012}. (Magnetization may change on longer
time scales due to, for example, dipolar interactions, atom loss, or
spurious $p$-wave scattering with high-temperature atoms.)
We have determined the structure and stability of
singular and non-singular vortices by numerically relaxing the energy
of trial wave functions, explicitly imposing conservation of
magnetization throughout the relaxation procedure. 
In order to describe vortex states that mix FM and polar phases, we
have analytically constructed spinor wave functions for vortices that smoothly
interpolate between the two manifolds.  In particular we have
explicitly derived solutions that describe the filled cores of
singular vortices.  Such filling of the vortex core occurs very
generally in the spinor BECs as result of energetic competition
between characteristic length scales as the energy relaxes. We have
also provided analytic construction of solutions that represent
composite defects where both polar and FM limits
correspond to vortex states.  These exhibit different vortex topology on small
and large distances from the vortex line. 
An example of a stable composite defect
is a FM coreless vortex in the
polar interaction regime~\cite{lovegrove_prl_2014}.
In addition to composite defects
occurring purely as a result of conservation of magnetization, we have
suggested how vortices with a concentric hierarchy of cores may be
stabilized in experiment.

\begin{acknowledgments}
  We acknowledge financial support from the EPSRC and the Leverhulme
  Trust, as well as the use of the IRIDIS High Performance
  Computing Facility at the University
  of Southampton.
\end{acknowledgments}

\appendix*

\section{Analytic wave functions for FM vortices with polar
  half-quantum vortex core}
\label{app:further-examples}

Just as the singly quantized polar vortex may form inside the core of a singular
FM vortex  (or vice versa), a polar half-quantum vortex
may form inside the singular FM vortex core (and vice versa), as
demonstrated in section~\ref{sec:composite-core}. Here we
construct an
explicit spinor for this composite vortex state by requiring
$\tau=-\gamma=\varphi/2$ and $\alpha=0$. Taking constant $\beta=\beta_0$,
Eq.~\eqref{eq:general} then becomes
\begin{equation}
  \label{eq:fm-hq1}
  \zeta= \frac{\mathcal{E}}{2}
    \threevec{\sqrt{2}\left(
               \mathcal{E}^{-1}\sin^2\frac{\beta_0}{2}D_-
	       -\mathcal{E}\cos^2\frac{\beta_0}{2}D_+\right)}
	     {-\sin\beta_0\left(\mathcal{E}^{-1}D_-
	       +\mathcal{E} D_+  \right)}
	     {\sqrt{2}\left(
	       \mathcal{E}^{-1}\cos^2\frac{\beta_0}{2}D_-
	       - \mathcal{E}\sin^2\frac{\beta_0}{2}D_+\right)},
\end{equation}
where $\mathcal{E}=\exp(i\varphi/2)$.
In the $F=1$ limit we recover the singular FM
vortex~\eqref{eq:fmsingular} with uniform spin profile
$\expF=\sin\beta_0\xhat+\cos\beta_0\zhat$.
The $F=0$ limit, on the other hand, is a half-quantum vortex with
$\nematic =
\cos\beta_0\cos(\varphi/2)\xhat-\sin(\varphi/2)\yhat -
\sin\beta_0\cos(\varphi/2)\zhat$
exhibiting the characteristic $\pi$ winding as the vortex line is encircled.
Hence, the half-quantum vortex forms the core of the FM singular
vortex when $F(\rho)$ increases from $F=0$ to $F=1$ with the radial distance.
Correspondingly, the circulation
$\nu=h(1+F)/2m$ interpolates smoothly between the inner half-quantum
of circulation and the outer singly quantized FM vortex.

The FM order parameter also allows the formation of a singular vortex
on the form of Eq.~\eqref{eq:spinvortex}, which is topologically
equivalent to Eq.~\eqref{eq:fmsingular}, and the one may be deformed
into the other by local spin rotations. We should therefore expect
also the core of Eq.~\eqref{eq:spinvortex} to be able to host a polar
half-quantum vortex. We find an expression for this composite vortex state
by taking
$\tau=\gamma=\varphi/2$, $\alpha=\varphi$. Then
\begin{equation}
  \label{eq:fm-hq2}
  \zeta= \frac{\mathcal{E}}{2}
    \threevec{\sqrt{2}\left(
               \mathcal{E}^{-1}\sin^2\frac{\beta(\rho)}{2}D_-
	       -\mathcal{E}^{-3}\cos^2\frac{\beta(\rho)}{2}D_+
	       \right)}
	     {-\sin\beta(\rho)\left(\mathcal{E} D_-
	       +\mathcal{E}^{-1}D_+ \right)}
	     {\sqrt{2}\left(
	       \mathcal{E}^3\cos^2\frac{\beta(\rho)}{2}D_-
	       - \mathcal{E}\sin^2\frac{\beta(\rho)}{2}D_+
	       \right)},
\end{equation}
The $F=1$ limit represents the FM singular
vortex~\eqref{eq:spinvortex}, exhibiting the radial spin disgyration
$\expF=\sin\beta(\rho)\rhohat+\cos\beta(\rho)\zhat$, by construction.
The angle $\beta(\rho)$
increases from $\pi/2$ to $\pi$ as a function of
radius.
In the $F=0$ limit, this spinor does indeed represent a
half-quantum vortex with $\tau$ winding by $\pi$.
The winding of $\nematic$ depends on $\beta$, resulting
in a nematic axis profile of
$\nematic=(\cos\varphi\cos\varphi/2 \cos\beta(\rho) - \sin\varphi
\sin\varphi/2)\xhat
+(\sin\varphi\cos\varphi/2 \cos\beta(\rho) + \cos\varphi \sin\varphi/2)\yhat
-\cos\varphi/2 \sin\beta(\rho)\zhat$.
Again the composite defect is formed as $F(\rho)$ is allowed
to vary such that $F(\rho\to0)=0$ and $F(\rho)=1$ at large $\rho$, and
the circulation 
$\nu=h(1-F-2F\cos\beta(\rho))/2m$ interpolates from the inner
half-quantum vortex of the polar phase to the non-quantized circulation
of the outer radial spin disgyration.

%

\end{document}